\documentstyle[preprint,floats,eqsecnum,tighten,prd,aps]{revtex}
\input epsf.tex
\def\DESepsf(#1 width #2){\epsfxsize=#2 \epsfbox{#1}}
\begin{document}
\preprint{\vbox{
\hbox{OITS 629}    
\hbox{SLAC-PUB-7480} 
\hbox{June 1997}
} }
\draft

\title{Virtual photon scattering at high energies \\
as a probe of the short distance pomeron}
\author{S.J.\ Brodsky$^{a}$, F.\ Hautmann$^{b}$ and D.E.\ Soper$^{b}$}
\address{$^{a}$ Stanford Linear Accelerator Center, 
Stanford University, Stanford, CA 94309}
\address{$^{b}$ Institute of Theoretical Science, 
University of Oregon, Eugene, OR 97403}
\maketitle

\begin{abstract}
Perturbative QCD predicts the behavior of scattering at high energies
and fixed (sufficiently large) transferred momenta in terms of the
BFKL pomeron (or short distance pomeron). We study the prospects
for testing these predictions in two-photon processes  at LEP200 and
a possible future $e^\pm e^-$ collider. We argue that the total cross
section for scattering two photons sufficiently far off shell
provides a clean probe of BFKL dynamics. The photons act as color
dipoles with small transverse size, so that the QCD interactions can
be treated perturbatively. We analyze the properties of the QCD result
and the possibility of testing them experimentally. We give an
estimate of the rates expected and discuss the uncertainties of these
results associated with the accuracy of the present theoretical
calculations. 
\end{abstract}

\pacs{}



\section{Introduction}
\label{sec:intro}

The behavior of scattering in the limit of high energy and fixed
momentum transfer is described in QCD, at least for situations in
which perturbation theory applies, by the
Balitskii-Fadin-Kuraev-Lipatov (BFKL) pomeron~\cite{BFKL} (or short
distance pomeron). Attempts to test experimentally this sector of QCD
have started in the last few years, mainly based on measurements of
deeply inelastic events at low values of the Bjorken variable $x$ in
lepton-hadron scattering~\cite{abra} and jet production at large
rapidity separations in hadron-hadron collisions~\cite{jetrap}. In
this paper we study the possibilities for investigating QCD pomeron
effects in a different context, namely in photon-photon scattering at
$e^{+} \, e^{-}$ colliders, where the photons are produced from the
lepton beams by bremsstrahlung. The results of this paper also apply
to $e^{-} \, e^{-}$ or $\mu^{\pm} \, \mu^{-}$ colliders. Some aspects
of this study have been presented in
Refs.~\cite{talksjb,talkfh,earlier}.

The quantity we consider is the total cross section for off-shell
photon scattering at high energy. This can be measured in $e^{+}
\, e^{-}$ collisions in which both the outgoing leptons are tagged.
This cross section presents some theoretical advantages as a probe of
QCD pomeron dynamics compared to the structure functions for deeply
inelastic scattering off a proton (see for instance the discussion in
Ref.~\cite{abra}) or a (quasi)-real photon (see for example
Ref.~\cite{gagaphy}), essentially because it does not involve a
non-perturbative target. Unlike protons or quasi-real photons,
virtual photon states can be described through perturbative wave
functions. In some respects the off-shell photon cross section
presents analogies with the process of scattering of two quarkonia
(or ``onia''), which has been proposed as a gedanken experiment to
investigate the high energy regime in QCD~\cite{onium}. In this case,
non-perturbative effects are suppressed by the smallness of the onium
radius. In the case of virtual photons the size of the wave function
is controlled by the photon virtuality instead of the heavy quark
mass. It is an interesting feature of investigations at $e^{+} e^{-}$
colliders that this size can be tuned by measuring the momenta of the
outgoing leptons.

On the other hand, such experimental studies may prove to be difficult
due to the smallness of the available rates. As we shall see, for
large photon virtualities $Q^2$ the cross section falls off like $1 /
Q^2$. An estimate of the number of events that one may expect to be
available for such studies at LEP200 and a future linear $e^+e^-$
collider will be provided later in the paper.

There have recently been other investigations of the high energy
regime in the context of photon-photon scattering.
Balitskii~\cite{bali} has proposed an expansion of the scattering
amplitude in the high energy limit in terms of Wilson line operators.
This method provides an elegant reformulation of the BFKL problem, and
may prove to be useful to get beyond the leading logarithm
approximation. Bartels, De Roeck and Lotter~\cite{bdl} have evaluated
the photon-photon cross section at present and future $e^{+} e^{-}$
colliders. Their results are similar to those in
Refs.~\cite{talksjb,talkfh,earlier}. In Ref.~\cite{gagaphy} detailed
studies have been carried out for diffractive meson production and
photon structure functions at LEP200.

\begin{figure}[htb]
\centerline{ \DESepsf(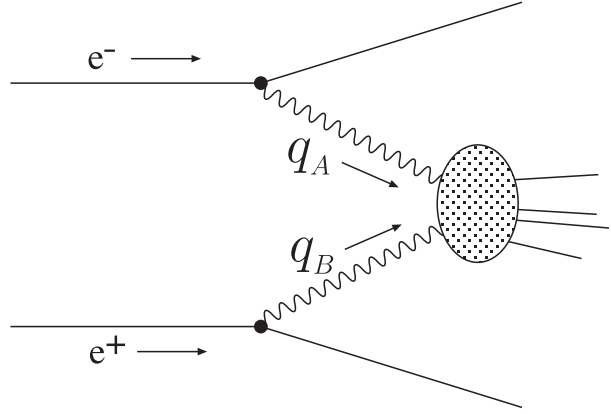 width 8 cm) }
\caption{ The photon-photon scattering process in $e^+ e^-$
collisions. }
\label{fee}
\end{figure}

To describe the electron-positron scattering process (Fig.~\ref{fee}),
we will parametrize the five-fold differential cross section as
\begin{equation}
\label{dsdphi}
{ { d \, \sigma^{(e^+ e^-)} } \over
{ d x_A \,
d x_B \,
d Q^2_A \,
d Q^2_B \, d \phi / (2 \, \pi) }} =
{ { d \, \sigma^{(e^+ e^-)} } \over
{ d x_A \,
d x_B \,
d Q^2_A \,
d Q^2_B }} \,
\left[ 1 + A_1 \, \cos{\phi} +
A_2 \, ( 2 \, \cos^2{\phi} - 1 ) \right]
\hspace*{0.8 cm} .
\end{equation}
Here we denote by $x_A, \, x_B$ the fractions of the longitudinal
momenta of the leptons A and B that are carried by the photons, by
$Q_A^2 = - q_A^\mu q_{A\,\mu}, \, Q_B^2 = - q_B^\mu q_{B\,\mu}$ the
photon virtualities, and by $\phi$ the angle between the lepton
scattering planes in a frame in which the photons are aligned with
the $z$ axis. The overall factor in the right hand side gives the
distribution averaged over the angle $\phi$, while $A_1$, $A_2$ are
the asymmetries.

We will start by considering the $\phi$-averaged cross section, and we
will express it in the equivalent photon approximation~\cite{budnev}
by folding the $\gamma^{*} \, \gamma^{*} $ cross section with the
flux of photons from each lepton. This flux is proportional to the
probability density for the splitting $e \to e \, \gamma$, and
depends on the photon polarization. We will thus write
\begin{eqnarray}
\label{epaave}
{ { Q^2_A \, Q^2_B \, d \, \sigma^{(e^+ e^-)} } \over
{ d x_A \,
d x_B \,
d Q^2_A \,
d Q^2_B }}
&=& \left( {\alpha \over {2 \, \pi}} \right)^2
\left\{
P_{\gamma/e^{+}}^{(T)}( x_A)
\, P_{\gamma/e^{-}}^{(T)}( x_B)\,
\sigma_{ {\gamma^{*}}{\gamma^{*}}}^{(TT)}
(x_A \, x_B \, s , Q_A^2, Q_B^2)
\right.
\nonumber\\
&&+
P_{\gamma/e^{+}}^{(T)}( x_A)
\, P_{\gamma/e^{-}}^{(L)}( x_B)\,
\sigma_{ {\gamma^{*}}{\gamma^{*}}}^{(TL)}
(x_A \, x_B \, s , Q_A^2, Q_B^2)
\nonumber\\
&&+
P_{\gamma/e^{+}}^{(L)}( x_A)
\, P_{\gamma/e^{-}}^{(T)}( x_B)\,
\sigma_{ {\gamma^{*}}{\gamma^{*}}}^{(LT)}
(x_A \, x_B \, s , Q_A^2, Q_B^2)
\nonumber\\
&&+
\left.
P_{\gamma/e^{+}}^{(L)}( x_A)
\, P_{\gamma/e^{-}}^{(L)}( x_B) \,
\sigma_{ {\gamma^{*}}{\gamma^{*}}}^{(LL)}
(x_A \, x_B \, s , Q_A^2, Q_B^2)
\right\}
\,\;\;\;,
\end{eqnarray}
where the transverse and longitudinal photon flux factors
$ P^{(T)}$, $ P^{(L)}$ are given by
\begin{equation}
\label{gammafluxtl}
P_{\gamma/e}^{(T)}( x) = {{1 + (1 - x)^2 } \over x} \hspace*{0.6 cm} ,
\hspace*{0.8 cm}
P_{\gamma/e}^{(L)}( x) = 2 \, {{1 - x } \over x} \hspace*{0.8 cm} ,
\end{equation}
and $\sigma_{ {\gamma^{*}}{\gamma^{*}}}^{(a \, b)}$, with
$a , b = T, L$, is the cross section for the scattering of two
photons with polarizations $a$ and $b$.

We will proceed in the following way. In Sec.~\ref{sec:notations} we
set our notations and describe the Born approximation to the
photon-photon cross section at high energy. In this section we
concentrate on the case of transversely polarized photons, that is,
the cross section $\sigma_{ {\gamma^{*}}{\gamma^{*}}}^{(TT)}$ in
Eq.~(\ref{epaave}) above. In Sec.~\ref{sec:polarization} we extend
these results to include the full polarization dependence and we
discuss the associated asymmetries. In Sec.~\ref{sec:summation} we
consider the summation of the leading logarithmic corrections to the
photon-photon cross section due to BFKL pomeron exchange, and
emphasize how the perturbative results depend on the total energy and
the photon virtualities. Secs.~\ref{sec:scales} and
\ref{sec:limitations} are devoted to discussing some of the
limitations of the treatment based on the BFKL equation. In
Sec.~\ref{sec:scales} we focus on the dependence of the cross section
on two mass scales (in the running coupling and in the high energy
logarithms), which are left undetermined in a leading log analysis.
In Sec.~\ref{sec:limitations} we consider the limitations of using the
BFKL approach that follow from the behavior at very large
$s$. Sec.~\ref{sec:virtuality} illustrates how to relate the summed
result for the photon-photon cross section to the small-$x$ behavior
of the photon deeply inelastic structure function. In
Sec.~\ref{sec:regge} we compare the QCD result with expectations
based on traditional Regge theory. In Sec.~\ref{sec:softandhard} we
consider the limit of low photon virtualities and discuss the region
of transition between hard and soft scattering. The rates at the
level of the $e^+ e^-$ cross section are examined in
Sec.~\ref{sec:numerical}. Some concluding remarks are given in
Sec.~\ref{sec:conclusions}. We collect in Appendix~\ref{app:born} the
details of the Born order calculation, and in
Appendix~\ref{app:quark} some formulas which are useful for comparing
the gluon-exchange and quark-exchange contributions to the
photon-photon cross section.

\section{Notations and lowest order calculation}
\label{sec:notations}

We consider the total cross section for the scattering of two
transversely polarized virtual (space-like) photons $\gamma^{*} (q_A)$
and $ \; \gamma^{*} (q_B)$, with virtualities $q_A^2 \equiv - Q^2_A$
and $q_B^2 \equiv - Q^2_B$, in the high energy region where the
center-of-mass energy $\sqrt{s} \equiv \sqrt{(q_A+q_B)^2}$ is much
larger than $Q_A$, $Q_B$. We also suppose that the photon
virtualities are in turn large with respect to the QCD scale
$\Lambda_{QCD}^2$, so that the process occurs at short distances
(much smaller than $\Lambda^{-1}_{QCD} \approx 1 \; {\mbox {fm}}$) and
QCD perturbation theory applies.

We work in a reference frame in which the incoming photons have zero
transverse momenta, and are boosted along the positive and negative
light-cone directions. For a four-momentum $p^\mu$, we define the
``$+$'' and ``$-$'' momentum components $p^{+}$, $p^{-}$ as
\begin{equation}
\label{pmcomps}
p^{\pm} = \left( p^{0} \pm p^{3} \right) / \sqrt{2} \;\;\;.
\nonumber
\end{equation}
The incoming photon momenta are parametrized as follows in a notation
where $q^\mu = (q^+,q^-,{\bf q}_T)$:
\begin{equation}
\label{qaqb}
q^{\mu}_A = \left( q^{+}_A, -{Q_A^2 \over {2 \; q^{+}_A}},
{\mbox {\bf 0}} \right) \;\;, \;\;\;
q^{\mu}_B = \left( -{Q_B^2 \over {2 \; q^{-}_B}}, q^{-}_B,
{\mbox {\bf 0}} \right) \;\;.
\end{equation}
Here the ``$+$'' and ``$-$'' components $q^{+}_A, q^{-}_B$
approximately build up the total energy $s$, and are much larger than
the initial virtualities
\begin{equation}
\label{qaplusqbminus}
2 \,
q^{+}_A \,
q^{-}_B \approx s \gg Q_A^2, Q_B^2
\;\;\;.
\end{equation}

\begin{figure}[htb]
\centerline{ \DESepsf(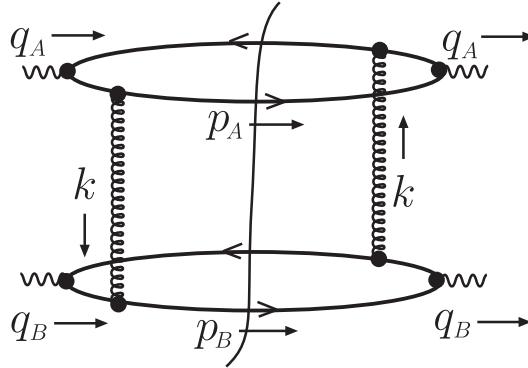 width 7 cm) }
\caption{ One of the two-gluon exchange graphs contributing to the
high energy $\gamma^* \gamma^*$ cross section in the Born
approximation. The gluons can be attached to the quark lines in $2^4$
different ways.}
\label{ftwoglu}
\end{figure}

The Born contribution to this cross section is given by the high
energy approximation to the reaction
\begin{equation}
\label{born}
\gamma^{*} (q_A) + \gamma^{*} (q_B) \to
q(p_A) + {\bar q}({\bar p}_A) +
q(p_B) + {\bar q}({\bar p}_B)
\;\;\;
\end{equation}
in lowest order perturbation theory. The corresponding diagrams
involve the exchange of two gluons between the two quark-antiquark
pairs~\cite{lownus}, and are exemplified in Fig.~\ref{ftwoglu}.

We parametrize the outgoing quark momenta as
\begin{equation}
\label{papb}
p^{\mu}_A = \left( z_A q^{+}_A,
z_A^{\prime} q^{-}_B,
{{\bf p}}_A \right) \;\;, \;\;\;
p^{\mu}_B = \left(
z_B^{\prime} q^{+}_A,
z_B q^{-}_B,
{{\bf p}}_B \right) \;\;,
\end{equation}
and denote by $k^{\mu}$ the exchanged gluon momentum.

In the high energy limit defined by Eq.~(\ref{qaplusqbminus}), the
kinematic region which dominates the integrations is the one in which
the transverse momenta flowing in the loops are of the order of the
initial virtualities, and the light-cone components of the exchanged
gluon are suppressed with respect to the transverse momenta by a
quantity of order $Q_A / \sqrt{s}$ or $Q_B / \sqrt{s}$, so that
$k^2 \simeq - {{\bf k}}^2$.

The squared amplitude for the graph in Fig.~\ref{ftwoglu}, integrated
over the final quark and antiquark phase space, for fixed transverse
photon polarizations $\varepsilon_A$, $\varepsilon_B$, is given by
\begin{eqnarray}
\label{m1}
&& {|{\cal M}|}^2 = \int \; {{d^4 \, k} \over
{(2 \, \pi)^4}} \;
{{d^4 \, p_A} \over {(2 \, \pi)^4}} \; {{d^4 \, p_B} \over
{(2 \, \pi)^4}}
\; {1 \over {(k^2)^2}}
\; 2 \, \pi \, \delta_{+}\!\left( p_A^2 \right) \,
\; 2 \, \pi \, \delta_{+}\!\left( p_B^2 \right) \, \sum_{a , b} \,
\\
&& \times \,
\; 2 \, \pi \, \delta_{+}\!\left( (q_A-p_A-k)^2 \right) \,
\; 2 \, \pi \, \delta_{+}\!\left( (q_B-p_B+k)^2 \right) \, \,
\nonumber \\
&& \times \,
{ {
{\rm Tr} \Bigl[ 
{\rlap/p}_A \, (i \, g_s t^r \gamma_\alpha) \, i \,
( {\rlap/p}_A + \rlap/k )  (i \, e_a \, e \, {\rlap/\varepsilon}_A
)  ({\rlap/q}_A - {\rlap/p}_A - \rlap/k ) 
( - i \, g_s t^s \gamma_\beta) ( - i) 
( {\rlap/p}_A - {\rlap/q}_A )  ( - i \, e_a \, e \,
{\rlap/\varepsilon}_A )
\Bigr]
} \over {
\left( (p_A + k)^2 + i \, \varepsilon \right)
\left( (p_A - q_A)^2 - i \, \varepsilon \right) } } \, \,
\nonumber \\
&& \times \,
{ {
{\rm Tr}\Bigl[ 
{\rlap/p}_B  (i g_s t^r \gamma^\alpha)\,  i\, 
( {\rlap/p}_B - \rlap/k ) \, (i \, e_b \, e  {\rlap/\varepsilon}_B
) \, ({\rlap/q}_B - {\rlap/p}_B + \rlap/k ) 
( - i  g_s t^s \gamma^\beta)  ( - i) 
( {\rlap/p}_B - {\rlap/q}_B )  ( - i e_b  e 
{\rlap/\varepsilon}_B )
\Bigr]
} \over {
\left( (p_B - k)^2 + i \, \varepsilon \right)
\left( (p_B - q_B)^2 - i \, \varepsilon \right) } } \;\;.
\nonumber
\end{eqnarray}
Here $e_a$, $e_b$ are the electric charges of the quarks in units of
$e = \sqrt{ 4 \, \pi \, \alpha}$, and the indices $a , b$ run over
the light quark flavors, $u , d , s$. (The flavors $c , b$ need a
separate treatment.) In the high energy approximation this amplitude
takes the form
\begin{eqnarray}
\label{m3}
&& {|{\cal M}|}^2 =
32 \, \alpha^2 \, \alpha_s^2 \,
\left( \sum_q e^2_q \right)^2
\, (2 \, \pi)^2 \, (2 \, s) \,
\int \; {{d^2 \, {\bf k}} \over {(2 \, \pi)^2}} \;
{{d^2 \, {{\bf p}}_A} \over {(2 \, \pi)^2}} \;
{{d^2 \, {{\bf p}}_B} \over {(2 \, \pi)^2}} \;
\int_0^1
{d \, z_A}
\int_0^1
{d \, z_B}
\, {1 \over {({{\bf k}}^2)^2} } \,
\nonumber
\\
&& \times \, { {
\left[ 4 z_A (1 - z_A)
{\bf{\varepsilon}}_A \cdot {{\bf p}}_A \;
{\bf{\varepsilon}}_A \cdot ({{\bf p}}_A + {\bf k}) -
{{\bf p}}_A \cdot ({{\bf p}}_A + {\bf k}) \right]
} \over {
\left[ z_A \, (1-z_A) \, Q^2_A
+ ({{\bf p}}_A + {{\bf k}})^2 \right] \,
\left[ z_A \, (1-z_A) \, Q^2_A
+ {{\bf p}}_A^2 \right] \,
} } \,
\nonumber \\
&& \times \, { {
\left[ 4 z_B (1 - z_B)
{\bf{\varepsilon}}_B \cdot {{\bf p}}_B \;
{\bf{\varepsilon}}_B \cdot ({{\bf p}}_B - {\bf k}) -
{{\bf p}}_B \cdot ({{\bf p}}_B - {\bf k}) \right]
} \over {
\left[ z_B \, (1-z_B) \, Q^2_B
+ ({{\bf p}}_B - {{\bf k}})^2 \right] \,
\left[ z_B \, (1-z_B) \, Q^2_B
+ {{\bf p}}_B^2 \right] \,
} } \;\;\;\;\;\;\;\;.
\end{eqnarray}
See Appendix A for details of this calculation.

The amplitudes for the other graphs, in which the gluons are
connected to different fermion lines, can be derived from
Eq.~(\ref{m3}) by using the replacements
\begin{equation}
\label{repl1}
\left[ z_A \, (1-z_A) \, Q^2_A
+ ({{\bf p}}_A + {{\bf k}})^2 \right] \to
{{1-z_A} \over {z_A}} \, \left[ z_A \, (1-z_A) \, Q^2_A
+ {{\bf p}}_A^2 \right] \,
\end{equation}
in the denominator, and
\begin{equation}
\label{repl2}
({{\bf p}}_A + {{\bf k}}) \to
- {{1-z_A} \over { z_A}} \,
{{\bf p}}_A \,
\end{equation}
in the numerator. (Analogous replacements hold for the momentum
components of the quark $p_B$). In addition, the contribution from
the graphs in which the quark and antiquark lines are interchanged
can be obtained by symmetrizing the above expressions with respect to
$ {{\bf p}}_A \to {{\bf p}}_A + {{\bf k}} $ (and, analogously,
$ {{\bf p}}_B \to {{\bf p}}_B - {{\bf k}} $).

We add the graphs, and divide by $ 2 \, s$ to form the $\gamma^* \,
\gamma^*$ cross section
\begin{equation}
\label{sigma}
\sigma^{(0)}_{ {\gamma^{*}} {\gamma^{*}} } = {1 \over {2 \, s}} \,
{|{\cal M}|}^2 \;\;\;\;.
\end{equation}
We find
\begin{equation}
\label{convsigma}
\sigma^{(0)}_{ {\gamma^{*}} {\gamma^{*}} } (s, Q_A^2, Q_B^2,
{\bf{\varepsilon}}_A, {\bf{\varepsilon}}_B)
=
{1 \over {2 \, \pi}}
\,
\int \, {{d^2 \, {\bf k}} \over { \pi}} \,
{1 \over {({{\bf k}}^2)^2} } \,
G({\bf k}; Q^2_A, {\bf{\varepsilon}}_A) \,
G( - {\bf k}; Q^2_B, {\bf{\varepsilon}}_B) \;\;\;\;.
\end{equation}
The factors $1 / \left( {{\bf k}}^2 \right)^2$ in this formula come
from the gluon propagators. These factors multiply functions $G ({\bf
k}; Q^2, {\bf{\varepsilon}})$, which describe the coupling of the
exchanged gluon to the quark-antiquark system created by the virtual
photon with virtuality $Q^2$ and transverse polarization
${\bf{\varepsilon}}$. The explicit expression for $G$ is
\begin{eqnarray}
\label{capitalg}
&~& G({\bf k}; Q^2, {\bf{\varepsilon}}) =
4 \, \alpha \, \alpha_s \, \left( \sum_q e^2_q \right) \,
\int \; {{d^2 \, {{\bf p}}} \over {\pi}} \;
\int_0^1 \,
{d \, z}
\,
\\
&~& \times \, \left\{
{ {
\left[ {{\bf p}}^2 - 4 z (1 - z)
({\bf{\varepsilon}} \cdot {{\bf p}} )^2\;
 \right]
} \over {
\left[ {{\bf p}}^2 + z \, (1-z) \, Q^2
\right]^2
} }
- { {
\left[ {{\bf p}} \cdot ({{\bf p}} + {\bf k})
- 4 z (1 - z)
{\bf{\varepsilon}} \cdot {{\bf p}} \;
{\bf{\varepsilon}} \cdot ({{\bf p}} + {\bf k})
\right]
} \over {
\left[
({{\bf p}} + {{\bf k}})^2 + z \, (1-z) \, Q^2 \right] \,
\left[
{{\bf p}}^2 + z \, (1-z) \, Q^2 \right]
} }
\right\} \;\;\;\;.
\nonumber
\end{eqnarray}
The functions $G$ can be thought of as ``color functions'' of the
virtual photon since they describe the color flow in the $q \bar q$
states. From the point of view of light-cone perturbation
theory~\cite{lcqed}, they correspond to the coupling of the
null-plane photon wave function to gluons.

In the remainder of this section we discuss the case of the average
over the two transverse photon polarizations. The detailed
polarization dependence of the color functions $G$ and the associated
polarization asymmetry in the cross section are treated in
Sec.~\ref{sec:polarization}. By taking the polarization average
\begin{equation}
\label{avep}
{1 \over 2} \, \sum_{\lambda} {\bf{\varepsilon}}_i^{(\lambda)}
{\bf{\varepsilon}}_j^{(\lambda)}
\to {1 \over 2} \, \delta_{\, i \, j \,} \hspace*{0.6 cm} ,
\nonumber\\
\end{equation}
we define the function $G_1$
\begin{equation}
\label{avg}
G_1 ({{\bf k}}^2 / Q^2) =
{1 \over 2} \, \sum_{\lambda} G({\bf k}; Q^2,
{\bf{\varepsilon}}^{(\lambda)})
\hspace*{0.6 cm} .
\end{equation}
We find
\begin{eqnarray}
\label{gi1funct}
&~& G_1({{\bf k}}^2 / Q^2) =
4 \, \alpha \, \alpha_s \, \left( \sum_q e^2_q \right) \,
\int \; {{d^2 \, {{\bf p}}} \over {\pi}} \;
\int_0^1 \,
{d \, z} \left[ z^2 + (1 - z)^2 \right]
\,
\nonumber \\
&~& \times \, \left[
{ {
{{\bf p}}^2
} \over {
\left[ {{\bf p}}^2 + z \, (1-z) \, Q^2
\right]^2
} } -
{ {
{{\bf p}} \cdot ({{\bf p}} + {\bf k})
} \over {
\left[
({{\bf p}} + {{\bf k}})^2 + z \, (1-z) \, Q^2 \right] \,
\left[ {{\bf p}}^2 + z \, (1-z) \, Q^2
\right]
} }
\right] \;\;.
\end{eqnarray}
Each one of the two ${{\bf p}} $-dependent terms in the integrand of
Eq.~(\ref{gi1funct}) would lead to an ultraviolet divergent integral,
but the divergence cancels in the sum, illustrating that the gluon
does not couple to the color singlet $q {\bar q}$ system in the limit
${{\bf p}}^2 \to \infty$. The photon virtuality $Q^2$ regularizes the
denominators in the infrared region, ${{\bf p}}^2 \to 0$. In the
limit of small $Q^2$, the probability density $P_{q / \gamma}(z) =
\left[ z^2 + (1 - z)^2 \right] / 2$ for the splitting of a
transversely polarized photon into a quark-antiquark pair
$(\gamma \to q {\bar q} )$ correctly factors out in front of the
logarithmic singularity $ d {{\bf p}}^2 / {{\bf p}}^2 $
associated with the region of strong ordering $ Q^2 \ll {{\bf p}}^2
\ll {{\bf k}}^2$.

By introducing the Feynman parametrization
\begin{eqnarray}
\label{feynpar}
&& {1 \over {
\left[
({{\bf p}} + {{\bf k}})^2 + z \, (1-z) \, Q^2 \right] \,
\left[ {{\bf p}}^2 + z \, (1-z) \, Q^2
\right]^2} }
\nonumber \\
&& \hspace*{2 cm}
\, = \int^1_0 \, d \lambda \,
{ {2 \,(1 - \lambda)} \over {
\left[
({{\bf p}} + \lambda \, {{\bf k}})^2 + z \, (1-z) \, Q^2 +
\lambda \, (1 - \lambda) {{\bf k}}^2 \right]^3 } }
\end{eqnarray}
and carrying out the integration over the shifted transverse momentum
variable $ {{\bf p}}^{\prime} = {{\bf p}} + \lambda \, {{\bf k}} $ in
Eq.~(\ref{gi1funct}), one can obtain the following useful
representation of the function $G_1({{\bf k}}^2 / Q^2)$ as an
integral over two dimensionless variables:
\begin{equation}
\label{twodimintg}
G_1({{\bf k}}^2 / Q^2) =
2 \, \alpha \, \alpha_s \, 
\left( \sum_q e^2_q \right) \, {{\bf k}}^2 \,
\int_0^1 \,
{d \, z} \, \int_0^1 \,
{d \, \lambda} \,
{ {
\left[ \lambda^2 + (1 - \lambda)^2 \right] \,
\left[ z^2 + (1 - z)^2 \right]
} \over {
\lambda \, (1 - \lambda) \,{{\bf k}}^2 + z \, (1-z) \, Q^2
} } \hspace*{0.5 cm} .
\end{equation}
This representation explicitly shows that the distribution
$\left( {{\bf k}}^2 \right)^{-1} G_1({{\bf k}}^2 / Q^2)$
is symmetric under interchange of the
space-like boson virtualities ${{\bf k}}^2$ and $Q^2$. In the
configurations in which one of the virtualities is much smaller than
the other one, this distribution is logarithmically enhanced. More
precisely, we find
\begin{equation}
\label{ksmall}
{1 \over { {{\bf k}}^2 } } \, G_1({{{{\bf k}}^2} / {Q^2}})
\sim
4 \, \alpha \, \alpha_s \, \left( \sum_q e^2_q \right) \;
{2 \over 3} \,
{1 \over {Q^2}} \, \left[ \ln{{Q^2} \over {{{\bf k}}^2}} 
  + {\cal O} \left( 1 \right) \right]
\hspace*{0.5 cm} , \hspace*{0.8 cm} {{\bf k}}^2 \ll Q^2
 \hspace*{0.5 cm} ,
\end{equation}
\begin{equation}
\label{klarge}
{1 \over { {{\bf k}}^2 } } \, G_1({{{{\bf k}}^2} / {Q^2}})
\sim
4 \, \alpha \, \alpha_s \, \left( \sum_q e^2_q \right) \;
{2 \over 3} \,
{1 \over {{{\bf k}}^2}} \, 
\left[ \ln{  {{{\bf k}}^2} \over {Q^2} } 
  + {\cal O} \left( 1 \right) \right] 
\hspace*{0.5 cm} , \hspace*{0.8 cm} {{\bf k}}^2 \gg Q^2
\hspace*{0.5 cm} ,
\end{equation}
where the coefficient in front of the logarithm is given by the first
moment of the splitting density
\begin{equation}
\label{twothirds}
\int_0^1 \,
{d \, x} \left[ x^2 + (1 - x)^2 \right] = {2 \over 3}
\hspace*{1 cm} .
\end{equation}
The logarithmic behavior at small ${{\bf k}}^2$ comes from the region
${{\bf k}}^2 \ll {{\bf p}}^2 \ll Q^2$. Here the quark transverse
momentum is much smaller than the photon virtuality, and the quark
longitudinal momentum fraction is very small, $z \ll 1$ (or
$(1-z)\ll 1$). This region is sometimes referred to as the aligned-jet
region, and corresponds to configurations in which the $q {\bar q}$
system fluctuates to large sizes~\cite{aligned}. Note that while for
the case at hand of the $\gamma^* \, \gamma^*$ total cross section
this region contributes only a logarithmic enhancement, for the case
of non-inclusive processes, such as processes involving rapidity
gaps, this is expected to become the dominant
contribution~\cite{frastr}.

\begin{figure}[htb]
\centerline{ \DESepsf(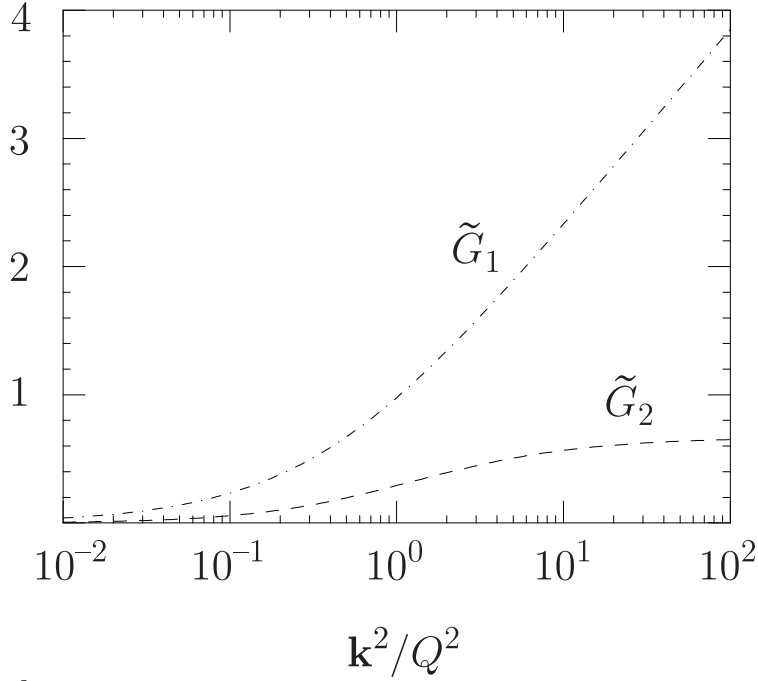 width 13 cm) }
\caption{ The ${{\bf k}}^2 / Q^2$ dependence of the color functions
$G_1$ (see Sec.~\protect\ref{sec:notations}) and $G_2$ (see
Sec.~\protect\ref{sec:polarization}). We plot the normalized
functions ${\tilde G} = G / (4 \, \alpha \, \alpha_s \,
\sum_q e^2_q)$. }
\label{glin}
\end{figure}

The integration over the parameter $\lambda$ in Eq.~(\ref{twodimintg})
can be explicitly performed. This yields
\begin{equation}
\label{etaxig}
G_1(\eta ) =
2 \, \alpha \, \alpha_s \, \left( \sum_q e^2_q \right) \,
\int_0^1 \, {{d \, \xi } \over { \sqrt{1 - \xi}}} \,
\left( 1 - { \xi \over 2} \right) \, \left[
{ { \eta + {\xi / 2} } \over {\sqrt{ \eta \, (\eta + \xi)}}} \,
\ln{ \left( { { \sqrt{\eta + \xi} + \sqrt{\eta} } \over
{ \sqrt{\eta + \xi} - \sqrt{\eta} } } \right) } - 1 \right]
\, ,
\end{equation}
where we have introduced the variables
\begin{equation}
\label{etaxi}
\xi \equiv 4 \, z \, (1 - z) \hspace*{1 cm} , \hspace*{2 cm}
\eta \equiv {{\bf k}}^2 / Q^2 \hspace*{1 cm} .
\end{equation}
In Fig.~\ref{glin}, we report the result of the numerical evaluation
of the integral (\ref{etaxig}) by plotting the function $G_1$ versus
$\eta = {{\bf k}}^2 / Q^2$.

The unpolarized $ \gamma^{*} \, \gamma^{*} $ cross section to Born
order, ${\overline \sigma}^{(0)}$, can be obtained by inserting the
result for the function $G_1$ in the general formula
(\ref{convsigma}). It is convenient to use the representation
(\ref{twodimintg}) of $G_1$. By substituting this representation in
Eq.~(\ref{convsigma}), and carrying out the integration over the
gluon transverse momentum
${{\bf k}}$, we obtain
\begin{eqnarray}
\label{sigmarepr}
&& {\overline \sigma}^{(0)} (s, Q_A^2, Q_B^2)
= {{2 \, \alpha^2 \, \alpha_s^2 \, \left( \sum_q
e^2_q \right)^2} \over {\pi}}
\, \int_0^1 \,
{d \, z_A} \left[ z_A^2 + (1 - z_A)^2 \right]
\int_0^1 \,
{d \, \lambda_A} \left[ \lambda_A^2 + (1 - \lambda_A)^2 \right] \,
\nonumber \\
&& \times
\int_0^1 \,
{d \, z_B} \left[ z_B^2 + (1 - z_B)^2 \right]
\int_0^1 \,
{d \, \lambda_B} \left[ \lambda_B^2 + (1 - \lambda_B)^2 \right] \,
\nonumber \\
&& \times { {
\ln \left[
{Q_A^2 \, z_A \, (1 - z_A) \, \lambda_B \, (1 - \lambda_B) } /
\left( {Q_B^2 \, z_B \, (1 - z_B) \, \lambda_A \, (1 - \lambda_A) }
\right)
\right]
}
\over
{Q_A^2 \, z_A \, (1 - z_A) \, \lambda_B \, (1 - \lambda_B) -
Q_B^2 \, z_B \, (1 - z_B) \, \lambda_A \, (1 - \lambda_A) }}
\hspace*{0.5 cm} .
\end{eqnarray}

\begin{figure}[htb]
\centerline{ \DESepsf(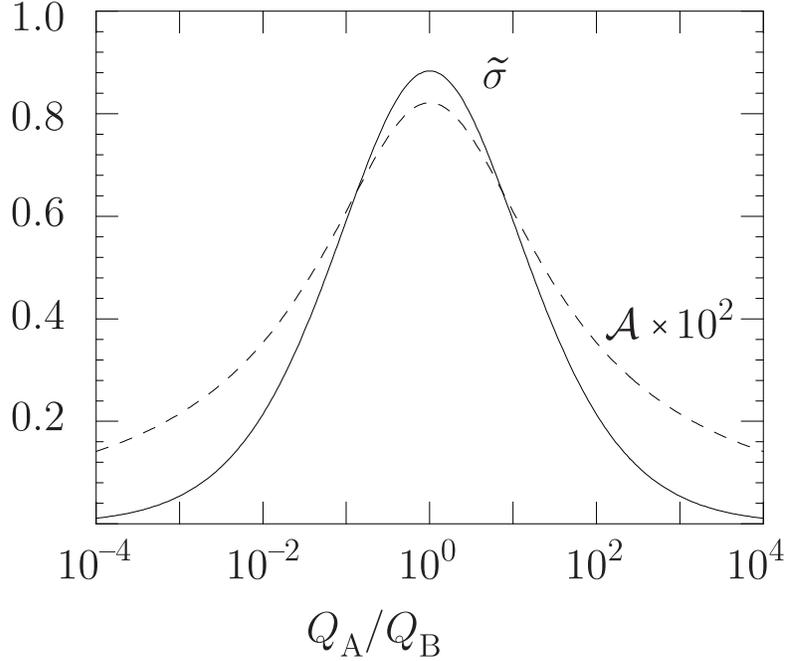 width 13 cm) }
\caption{ The high-energy $\gamma^* \gamma^*$ cross section in the Born
approximation as a function of the ratio $ r \equiv Q_A / Q_B$ between
the photon virtualities. The solid line is the rescaled cross section
$\tilde\sigma(r)$ defined in Eq.~(\protect\ref{tildesigmadef}).  The
dashed line is the polarization asymmetry discussed in
Sec.~\protect\ref{sec:polarization} (see
Eq.~(\protect\ref{convsiasymm})), multiplied by $10^2$. }
\label{s0}
\end{figure}

This formula provides an expression for the cross section in the
large $s$ limit in terms of dimensionless integrals, which allows us
to study the dependence on the energy and mass scales. To this order
in perturbation theory the cross section has a constant behavior with
the energy $s$. The cross section depends on the mass scales
$Q_A^2$ and $Q_B^2$ only. Factoring out an overall scale factor
$1 / ( Q_A \, Q_B)$ in Eq.~(\ref{sigmarepr}), we are left
with a function $\tilde\sigma(r)$ of the ratio $r \equiv Q_A / Q_B$:
\begin{equation}
\bar\sigma^{(0)} = 16\alpha^2\alpha_s^2 \left({\sum_q e_q^2}\right)^2
{\tilde\sigma(r) \over Q_A Q_B}.
\label{tildesigmadef}
\end{equation}
The function $\tilde\sigma(r)$ can be computed by performing
numerically the integrations over the dimensionless variables $z$'s and
$\lambda$'s. The result is plotted in Fig.~\ref{s0}.

Notice that the dependence of $\tilde\sigma(r)$ on $r$ is rather mild
when $r$ is near 1. Thus, in this region, $\tilde\sigma(r)$ in
Eq.~(\ref{tildesigmadef}) could be treated as a constant, so that
\begin{equation}
\label{mild}
{\overline {\sigma}}^{(0)}
\sim 1 / \left( Q_A \, Q_B \right)
\hspace*{0.5 cm} ,
\hspace*{0.8 cm} r \approx 1 \hspace*{0.5 cm} .
\end{equation}
When, in contrast, the two virtualities are widely disparate from each
other, either $r \gg 1$ or $r \ll 1$, the function $\tilde\sigma(r)$
has a strong dependence on $r$, vanishing linearly (modulo logarithmic
enhancements) with either $r^{-1}$ or $r$, respectively. This provides
the overall change of scale in the cross section
\begin{equation}
\label{chofsca}
{\overline {\sigma}}^{(0)} \sim
1 / Q_A^{2} \hspace*{0.5 cm} ,
\hspace*{0.8 cm} Q_A^2 \gg Q_B^2 \hspace*{0.5 cm} ,
\end{equation}
\begin{equation}
\label{chofscb}
{\overline {\sigma}}^{(0)} \sim
1 / Q_B^{2} \hspace*{0.5 cm} ,
\hspace*{0.8 cm} Q_B^2 \gg Q_A^2 \hspace*{0.5 cm} .
\end{equation}
We will come back to this later on and examine the details of these
behaviors numerically (see Sec.~\ref{sec:virtuality}).

\section{Polarization dependence}
\label{sec:polarization}

We now study the dependence of the color functions $G$ and of the
photon-photon cross section on the polarization of the virtual photon.
Working in the frame defined by Eq.~(\ref{qaqb}), and denoting by $
u$, $v$ the lightlike unit vectors
\begin{equation}
\label{uandv}
u = \left( 1 , 0, {\mbox {\bf 0}} \right) \;\;, \;\;\;
v = \left( 0 , 1, {\mbox {\bf 0}} \right) \;\; ,
\end{equation}
we introduce the following decomposition of the polarization tensor
for the photon $q_A$:
\begin{equation}
\label{gmunu}
- g_{\mu \nu} = \sum_{i = 1 , 2} \varepsilon_{\mu}^{i} (q_A)
\varepsilon_{\nu}^{i} (q_A)
- { {Q_A^2} \over {(q_A \cdot v)^2} } v_\mu \, v_\nu
- {{ {q_A}_\mu \, v_\nu + v_\mu \, {q_A}_\nu } \over {q_A \cdot v} }
\;\; .
\end{equation}
The index $i$ in the first term in the right hand side of
Eq.~(\ref{gmunu}) runs over the transverse polarizations. We describe
these polarizations by using the linear basis
\begin{equation}
\label{linpol}
\varepsilon_{\mu}^{i} (q_A) = \delta_{\mu}^{i}
\hspace*{0.6 cm} , \hspace*{0.2 cm} i = 1, 2 \hspace*{0.6 cm} .
\end{equation}
From the second term in the right hand side of Eq.~(\ref{gmunu}), we
define the longitudinal polarization vector as
\begin{equation}
\label{epslong}
\varepsilon_{\mu}^{L} (q_A) = { {Q_A} \over {q_A \cdot v} } v_\mu
\;\;.
\end{equation}
The last term in Eq.~(\ref{gmunu}) does not contribute because of
current conservation. We thus have
\begin{equation}
\label{gmunubis}
- g_{\mu \nu} = \sum_{i = 1 , 2} \varepsilon_{\mu}^{i} (q_A)
\, \varepsilon_{\nu}^{i} (q_A)
- \varepsilon_{\mu}^{L} (q_A) \,
\varepsilon_{\nu}^{L} (q_A)
+ \{ gauge \;\; terms \} \hspace*{0.8 cm} .
\end{equation}
Formulas analogous to Eqs.~(\ref{gmunu})-(\ref{gmunubis}),
but with $v^\mu$ replaced by $u^\mu$, hold for the photon $q_B$.

Let us first consider the case of transverse polarizations. It is
convenient to introduce the tensor amplitude ${\cal G}^{\mu \, \nu}$
by rewriting Eq.~(\ref{capitalg}) in the form
\begin{equation}
\label{curlyg}
G({\bf k}; Q^2, {\bf{\varepsilon}}) = 
{\cal G}^{\, \mu \, \nu \,} ({\bf k};
Q^2) \;\; {\bf{\varepsilon}}_\mu \; {\bf{\varepsilon}}_{\nu}
\hspace*{0.6 cm} .
\end{equation}
The explicit expression for ${\cal G}^{\mu \, \nu}$ can be found in
Appendix A. For polarization indices $a , a^{\prime} = 1 , 2 , L$
we also introduce the notation
\begin{equation}
\label{quadrep}
{\cal G}^{\, \mu \, \nu \,} \;\;
{\bf{\varepsilon}}_{\mu}^{a} \;
{\bf{\varepsilon}}_{\nu}^{a^{\prime}} = G^{\,a \, a^{\prime} \,} \;\;
\hspace*{0.6 cm} .
\end{equation}
We can parametrize the transverse components of the tensor $\; G^{\,
a \, a^{\prime} \,} \; $ in terms of two scalar functions $\; G_1 \;
$ and $\; G_2$:
\begin{equation}
\label{gij}
G^{\, i \, i^{\prime} \, } ({\bf k}; Q^2) = \delta^{\, i \, i^{\prime} \,} \,
G_1 ({{\bf k}}^2 / Q^2)
-
\left( {{{{\bf k}}^i {{\bf k}}^{i^{\prime}} } \over {{{\bf k}}^2}} - {1 \over 2} \,
\delta^{\, i \, i^{\prime} \,} \right)
G_2 ({{\bf k}}^2 / Q^2)
\hspace*{0.4 cm} ,
\hspace*{0.6 cm} i , i^{\prime} = 1 , 2 \hspace*{0.6 cm} .
\end{equation}
The function $G_1$ represents the unpolarized color function, which
we have discussed in the previous section
(Eqs.~(\ref{avg}), (\ref{gi1funct})). The function $\; G_2$ carries
the information on the polarization dependence. Its explicit
expression is
\begin{eqnarray}
\label{g21}
&~& G_2({{\bf k}}^2 / Q^2) =
4 \, \alpha \, \alpha_s \, \left( \sum_q e^2_q \right) \,
\int \; {{d^2 \, {{\bf p}}} \over {\pi}} \;
\int_0^1 \,
{d \, z} \, 4 \, z \, (1 - z)
\,
\nonumber \\
&~& \times \, \left[ -
{ {
{{\bf p}}^2 - 2 \, ({{\bf p}} \cdot {\bf k})^2 / {{\bf k}}^2
} \over {
\left[ {{\bf p}}^2 + z \, (1-z) \, Q^2
\right]^2
} } +
{ {
{{\bf p}} \cdot ({{\bf p}} + {\bf k}) - 2 \, ({{\bf p}} 
+ {\bf k}) \cdot {\bf k} \;
{{\bf p}} \cdot {\bf k} / {{\bf k}}^2
} \over {
\left[
({{\bf p}} + {{\bf k}})^2 + z \, (1-z) \, Q^2 \right] \,
\left[ {{\bf p}}^2 + z \, (1-z) \, Q^2
\right]
} }
\right] \;\;.
\end{eqnarray}
The integrals in Eq.~(\ref{g21}) can be handled using the same
procedure as in the unpolarized case:
\begin{eqnarray}
\label{g22}
G_2({{\bf k}}^2 / Q^2) &=&
16 \, \alpha \, \alpha_s \, \left( \sum_q e^2_q \right) \, 
{{\bf k}}^2
\int_0^1 \,
{d \, z} \,
\int_0^1 \,
{d \, \lambda} \, \,
{ {
 \lambda \, (1 - \lambda) \, z \, (1 - z)
} \over {
\lambda \, (1 - \lambda) \,{{\bf k}}^2 + z \, (1-z) \, Q^2
} }
\\
&=&
2 \, \alpha \, \alpha_s \, \left( \sum_q e^2_q \right) \,
\int_0^1 \, {{d \, \xi } \over { \sqrt{1 - \xi}}} \,
\xi \, \left[ 1 -
{ { {\xi / 2} } \over {\sqrt{ \eta \, (\eta + \xi)}}} \,
\ln{ \left( { { \sqrt{\eta + \xi} + \sqrt{\eta} } \over
{ \sqrt{\eta + \xi} - \sqrt{\eta} } } \right) } \right]
\, ,
\nonumber
\end{eqnarray}
where we have used the variables defined in Eq.~(\ref{etaxi}).

Correspondingly, the cross section for scattering transversely
polarized photons, Eq.(\ref{convsigma}), can be decomposed as
\begin{equation}
\label{unplusass}
\sigma_{ {\gamma^{*}} {\gamma^{*}} } (s, Q_A^2, Q_B^2,
{\bf{\varepsilon}}_A, {\bf{\varepsilon}}_B) =
{\overline {\sigma}} (s, Q_A^2, Q_B^2) \left[ 1
+ \,
\left( 2 \, ({\bf{\varepsilon}}_A \cdot {\bf{\varepsilon}}_B)^2 - 1
\right) \, {\cal A} (s, Q_A^2, Q_B^2) \right]
\hspace*{0.6 cm} .
\end{equation}
The polarization average ${\overline {\sigma}}$ has been given to
order $\alpha_s^2$ in the previous section (Eq.~(\ref{sigmarepr})),
and the asymmetry ${\cal A}$ to the same order $\alpha_s^2$ is given
in terms of the color function $G_2$ as
\begin{equation}
\label{convsiasymm}
{\cal A} (s, Q_A^2, Q_B^2) =
{ 1 \over { {\overline {\sigma}} (s, Q_A^2, Q_B^2) }} \,
{1 \over {16 \, \pi}} \,
\int \, {{d^2 \, {\bf k}} \over { \pi}} \,
{1 \over {({{\bf k}}^2)^2} } \,
G_2({\bf k}^2/ Q^2_A) \,
G_2({\bf k}^2/ Q^2_B) \;\;\;\;.
\end{equation}

Numerical results are reported in Figs.~\ref{glin} and \ref{s0},
where we plot $G_2$ versus $\eta \equiv {{\bf k}}^2/Q^2$, and show the
dependence of ${\cal A}$ on the incoming photon virtualities. Unlike
$G_1$, $G_2$ is not logarithmically enhanced in the regions
$ {{\bf k}}^2 \ll Q^2$, $ {{\bf k}}^2 \gg Q^2$. This is related to the fact
that the splitting function associated with $G_2$ has zeros at the
endpoints of the spectrum in the longitudinal momentum fraction,
$ z=0$ and $z=1$ (see Eq.~(\ref{g22})), whereas the unpolarized
splitting function goes to a finite constant. The asymmetry ${\cal A}$
associated with the photon-photon scattering process, calculated here
in the Born approximation, contributes to the asymmetry $A_2$ in
Eq.~(\ref{dsdphi}) at the level of the $e^+ \, e^-$ scattering.
Numerical results for the $e^+ \, e^-$ process will be given in
Sec.~\ref{sec:numerical}.

We now move on to the case of the longitudinal polarization. Let us
consider the longitudinal-longitudinal color function:
\begin{equation}
\label{longlong}
G^{\,L \, L \,} = {\cal G}^{\, \mu \, \nu \,} \;\;
{\bf{\varepsilon}}_{\mu}^{L} \; {\bf{\varepsilon}}_{\nu}^{L}
\;\;
\hspace*{0.6 cm} .
\end{equation}
Following the lines of the calculation described in detail for the
case of transverse photons, we find
\begin{eqnarray}
\label{gll1}
&~& G^{L L} ({{\bf k}}^2 / Q^2) =
4 \, \alpha \, \alpha_s \, \left( \sum_q e^2_q \right) \,
\int \; {{d^2 \, {{\bf p}}} \over {\pi}} \;
\int_0^1 \,
{d \, z} \, 4 \, z^2 \, (1 - z)^2
\,
\nonumber \\
&~& \times \, \left[
{ {
Q^2
} \over {
\left[ {{\bf p}}^2 + z \, (1-z) \, Q^2
\right]^2
} } -
{ {
Q^2
} \over {
\left[
({{\bf p}} + {{\bf k}})^2 + z \, (1-z) \, Q^2 \right] \,
\left[ {{\bf p}}^2 + z \, (1-z) \, Q^2
\right]
} }
\right] \;\;.
\end{eqnarray}
This contribution equals the color function $G_2$ given above. This
can be seen by introducing an integral over a Feynman parameter
$\lambda$, then integrating over the transverse momentum ${{\bf p}}$
in Eq.~(\ref{gll1}). We get
\begin{equation}
\label{gll2}
G^{L L}({{\bf k}}^2 / Q^2) =
16 \, \alpha \, \alpha_s \, \left( \sum_q e^2_q \right) \, {{\bf k}}^2
\int_0^1 \,
{d \, z} \,
\int_0^1 \,
{d \, \lambda} \, \,
{ {
 \lambda \, (1 - \lambda) \, z \, (1 - z)
} \over {
\lambda \, (1 - \lambda) \,{{\bf k}}^2 + z \, (1-z) \, Q^2
} } = G_2 \;\;.
\end{equation}
As noted above, $G_2$ has no logarithms at small ${\bf k}^2$. This
corresponds to the absence of aligned-jet terms for longitudinally
polarized photons~\cite{frastr}. Contributions from longitudinally
polarized photons enter the $e^+ \, e^-$ cross section (\ref{epaave}).
They will be included in the numerical estimates that we give in
Sec.~\ref{sec:numerical}.

Finally, we consider the interference contribution between
longitudinal and transverse polarizations:
\begin{equation}
\label{longtras}
G^{ \, i \, L \,} = G^{ L \, i } = {\cal G}^{\, \mu \, \nu \,} \;\;
{\bf{\varepsilon}}_{\mu}^{i} \; {\bf{\varepsilon}}_{\nu}^{L}
\hspace*{0.4 cm} ,
\hspace*{0.6 cm} i = 1 , 2 \hspace*{0.6 cm} .
\end{equation}
In the high energy approximation in which we are working, this
contribution vanishes. This can be seen explicitly by writing the
color function in the general form
\begin{equation}
\label{gtl}
G^{ i \, L } = G^{ L \, i } =
G_3 ( {{\bf k}}^2 / Q^2) \;\, {\mbox {\bf k}}^{i} 
/ |{\bf k} |
\hspace*{0.8 cm} ,
\end{equation}
and computing the invariant function $G_3$. We get
\begin{equation}
\label{g4}
G_3({{\bf k}}^2 / Q^2) =
16 \, \alpha \, \alpha_s \, \left( \sum_q e^2_q \right) \, 
| {{\bf k}} | \; Q \,
\int_0^1 \,
{d \, z} \,
\int_0^1 \,
{d \, \lambda} \, \,
{ {
 \lambda \, (1 - \lambda) \, z \, (1 - z) \, (1 - 2 \, z)
} \over {
\lambda \, (1 - \lambda) \,{{\bf k}}^2 + z \, (1-z) \, Q^2
} }
\equiv 0
\hspace*{0.5 cm} .
\end{equation}
Interference terms of the kind in Eq.~(\ref{gtl}) would give rise to
the asymmetry $A_1$ (see Eq.~(\ref{dsdphi})) at the level of the $e^+
\, e^-$ scattering process. We thus see that this asymmetry vanishes
in the high energy approximation.

\section{Summation of leading logarithms}
\label{sec:summation}

We have seen that the two-gluon exchange mechanism gives rise to a
constant $ \gamma^* \, \gamma^*$ total cross section at large $s$,
$\sigma^{(0)} (s, Q_A^2, Q_B^2) \sim \alpha^2 \,
\alpha_s^2 \, f (Q_A^2, Q_B^2)$. To higher orders in perturbation
theory, the iteration of gluon ladders in the $t$-channel promotes
this constant to logarithms, and the perturbative expansion of the
cross section at high energy has the form
\begin{equation}
\label{expansion}
\sigma_{\gamma^{*} \gamma^{*}} \sim
\sigma^{(0)} \,
\left[ 1 + \sum_{k=1}^{\infty} a_k \left( \alpha_s \,
L \right)^k
+ \dots \right] \hspace*{0.6 cm} , \hspace*{0.8 cm}
L = \ln (s/Q^2) \hspace*{0.6 cm} ,
\end{equation}
where $Q^2$ is a scale of the order of the initial photon
virtualities, the sum represents the series of the leading logarithms
to all orders in the strong coupling $\alpha_s$, and the dots stand
for non-leading terms.

To study the high energy behavior, it is convenient to analyze the
cross section in its Mellin-Fourier moments, defined by
\begin{equation}
\label{mellin}
\sigma (s, Q_A^2, Q_B^2) = \int_{a - i \infty}^{a + i \infty}
\, { { d \, N} \over { 2 \pi i} } \,
e^{ N \, L} \,
\sigma_N (Q_A^2, Q^2_B) \hspace*{0.5 cm} ,
\end{equation}
where the $N$-integral goes along a contour parallel to the imaginary
axis and to the right of any singularities in $\sigma_N$. We see from
this definition that a constant behavior of the cross section with
the energy $s$ is generated by a simple pole in the moments
$\sigma_N$ at $N = 0$, while powers of logarithms are generated by
multiple poles at $N = 0$. The inverse of (\ref{mellin}) is
\begin{equation}
\sigma_N(Q_A^2,Q_B^2) =
\int_0^\infty {d s \over s}\
\left(s \over Q^2\right)^{-N}
\sigma(s,Q_A^2,Q_B^2) \hspace*{0.5 cm} .
\end{equation}

To sum the leading logarithmic terms, the basic observation of
BFKL~\cite{BFKL} is that the logarithms arise when multiple soft
gluons are emitted into the final state. These gluons have transverse
momenta $k_\perp$ of the same order as $Q_A$ and $Q_B$ and have
strongly ordered rapidities, lying between the rapidities of the
quarks in photon $A$ and the quarks in photon $B$. Along with
emission of real gluons, one also includes the exchange of
corresponding virtual gluons.

\begin{figure}[htb]
\centerline{ \DESepsf(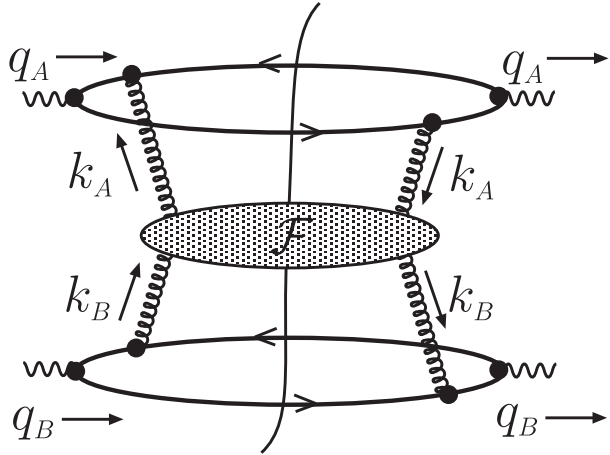 width 8 cm) }
\caption{Factorized structure of the virtual photon cross section
in the high energy limit. }
\label{fac}
\end{figure}

Fig.~\ref{fac} illustrates how the two gluon exchange graph of
Fig.~\ref{ftwoglu} is generalized to allow for multiple gluon
emission. The quarks comprising photon $A$ couple to a gluon with
momentum $k_A^\mu$, while the quarks comprising photon $B$ couple to
a gluon with momentum $k_B^\mu$. Gluons can be exchanged or emitted
into the final state inside the subgraph labeled $\cal F$. In fact,
the gluons that carry momenta $k_A^\mu$ and $k_B^\mu$ are, in
general, combinations of soft gluons that carry a net color octet
charge, that is, reggeized gluons~\cite{BFKL}. From a kinematic
viewpoint, we can treat these as being equivalent to ordinary
perturbative gluons. We consider the unpolarized $\gamma^*\gamma^*$
cross section and adopt the following notation for the diagram in
Fig.~\ref{fac}:
\begin{eqnarray}
\label{genfig}
\sigma(s,Q_A^2,Q_B^2)
&=&
\int { d^2 {\bf k}_A \over \pi {\bf k}_A^2}
\int_{-q_A^+}^0 dk_A^+\
J(-k_A^+/q_A^+,-{\bf k}_A;Q_A^2)
\\
&&\times
\int { d^2 {\bf k}_B \over \pi {\bf k}_B^2}
\int_0^{q_B^-} dk_B^-\
J(k_B^-/q_B^-,{\bf k}_B;Q_B^2)
\, {\cal G}(k_A^+,k_B^-,{\bf k}_A,{\bf k}_B) \;\;.
\nonumber
\end{eqnarray}
Here ${\cal G}$ represents the four-point Green function for gluons
$k_A$ and $k_B$, while the functions $J$ represent the quark loops.
We have made the following approximation. We note that the quark loop
for photon $A$ depends sensitively on the minus component, $k_A^-$,
of the momentum that enters the quark loop via gluon $A$. On the
other hand, the quarks in photon $B$ have very large minus components
of momenta. Furthermore, because of strong rapidity ordering, all the
other gluons inside of $\cal F$ have minus components of momenta that
are much larger than $k_A^-$. Thus we can neglect $k_A^-$ everywhere
except in the quark loop for photon $A$. Then we include the
integration over $k_A^-$ in the definition of
$J(-k_A^+/q_A^+,-{\bf k}_A;Q_A^2)$. Similarly we neglect $k_B^+$
everywhere except in the quark loop for photon $B$ and we include the
integration over $k_B^+$ in the definition of $J(k_B^-/q_B^-,{\bf
k}_B;Q_B^2)$.

Consider Eq.~(\ref{genfig}) specialized to the case of two gluon
exchange. The approximate gluon four-point function
${ \cal G}$ is trivial in this case:
\begin{equation}
{\cal G}_0(k_A^+,k_B^-,{\bf k}_A,{\bf k}_B)
= { 1 \over 2}\ \delta({\bf k}_A - {\bf k}_B )\
\delta(k_A^+)\ \delta(k_B^-) \hspace*{0.8 cm} .
\end{equation}
We identify
\begin{equation}
J(0,-{\bf k}_A;Q_A^2)
= G_1({\bf k}_A^2/Q_A^2)
\hspace*{0.5 cm} , \hspace*{0.8 cm}
J(0,{\bf k}_B;Q_B^2)
= G_1({\bf k}_B^2/Q_B^2) \hspace*{0.5 cm} .
\end{equation}
Then we recover the earlier result (\ref{convsigma}). 

For a generic term in $\cal G$, the transverse momenta
${\bf k}_A$ and ${\bf k}_B$ are independent variables, while the
factor $\delta(k_A^+)\, \delta(k_B^-)$ turns into
$1/(k_A^+ k_B^-)$. Taking moments with respect to $s$ and changing
integration variables to $\xi_A = -k_A^+/q_A^+$, $\xi_B =
k_B^-/q_B^-$ and $\lambda = \xi_A\xi_B s/Q^2$ gives the structure
\begin{eqnarray}
\sigma_N(Q_A^2,Q_B^2) &=&
\int { d^2 {\bf k}_A \over \pi {\bf k}_A^2}
\int_0^1 {d\xi_A\over \xi_A}\ \xi_A^{N}\
J(\xi_A,-{\bf k}_A;Q_A^2)
\\
&&\times
\int { d^2 {\bf k}_B \over \pi {\bf k}_B^2}
\int_0^1 {d\xi_B\over \xi_B}\ \xi_B^{N}\
J(\xi_B,{\bf k}_B;Q_B^2)
\int_0^\infty\ { d\lambda \over \lambda}\ \lambda^{-N}\
{\cal H}(\lambda Q^2,{\bf k}_A,{\bf k}_B) \;\;\;.
\nonumber
\end{eqnarray}
Having evaluated $\cal H$ at some fixed perturbative order, we are
interested in the poles of $\sigma_N$ at $N = 0$. There are poles
from the small $\xi_A$ and $\xi_B$ ends of the integrations over
$\xi_A$ and $\xi_B$. To evaluate $\sigma_N$ without losing any powers
of $1/N$, we approximate
\begin{eqnarray}
J(\xi_A,-{\bf k}_A;Q_A^2) &\to&
J(0,-{\bf k}_A;Q_A^2)
= G_1({\bf k}_A^2/Q_A^2) \;\;,
\nonumber\\
J(\xi_B,{\bf k}_B;Q_B^2) &\to&
J(0,{\bf k}_B;Q_B^2)
= G_1({\bf k}_B^2/Q_B^2) \;\;.
\end{eqnarray}
The pole associated with the integration over $\xi_A$ can be thought
of as arising from the integration over the rapidity of the final
state gluon with the largest rapidity. Almost all of the momentum
fraction $\xi_A$ is taken by this gluon. Since, in the leading
logarithmic approximation, this gluon has a rapidity that is much
less than that of the quarks in photon $A$, the momentum fraction
$\xi_A$ is negligible compared to 1. Similarly, the pole associated
with the integration over $\xi_B$ can be thought of as arising from
the integration over the rapidity of the final state gluon with the
most negative rapidity. The poles from the large $\lambda$ end of the
integration over the function $\cal H$ can be thought of as arising
from integrations over final state gluons with intermediate
rapidities.

The leading logarithmic result can thus be written in the form
\begin{equation}
\label{ktfac}
\sigma_N(Q_A^2,Q_B^2) =
\int { d^2 {\bf k}_A \over \pi {\bf k}_A^2}
\int { d^2 {\bf k}_B \over \pi {\bf k}_B^2}\
G_1({\bf k}_A^2/Q_A^2)\
{\cal F}_N({\bf k}_A^2, {\bf k}_B^2)\
G_1({\bf k}_A^2/Q_A^2) \;\;\;\;,
\end{equation}
where ${\cal F}_N({\bf k}_A^2, {\bf k}_B^2)$ is the BFKL function that
describes the interaction of gluons with fast moving colored systems
moving in opposite directions and obeys the BFKL
equation~\cite{BFKL}. It is normalized so that at order $\alpha_s^0$
we have
\begin{equation}
{\cal F}_N^{(0)}({\bf k}_A^2, {\bf k}_B^2)
= 
{ 1 \over {2 \, N}} \ {1 \over \pi}\,
\delta({\bf k}_A^2 - {\bf k}_B^2)
\;\;\;.
\end{equation}
The solution to the BFKL equation can be written to all orders in
$\alpha_s$ in the form
\begin{equation}
\label{gareprf}
{\cal F}_N ({{\bf k}}_A^2, {{\bf k}}_B^2) =
{ 1 \over { 2 \, \, \pi \, \sqrt{ {{\bf k}}^2_A \, {{\bf k}}^2_B} } } \,
\int_{1 / 2 - i \infty}^{1 / 2 + i \infty} \, {{d \, \gamma} \over
{2 \pi i}} \,
\left( {{ {{\bf k}}^2_A } \over { {{\bf k}}^2_B} } \right)^{\gamma - 1 / 2 }
\, {1 \over {N - {\bar \alpha_s}
\, \chi (\gamma) }} \hspace*{0.8 cm} ,
\end{equation}
where
\begin{equation}
\label{abar}
{\bar \alpha_s} = \alpha_s
\, C_A / \pi \hspace*{0.5 cm} , \hspace*{0.8 cm} C_A = 3
\hspace*{0.8 cm} .
\end{equation}
The function $\chi (\gamma)$ is determined by solving the eigenvalue
problem for the BFKL kernel and is given by
\begin{equation}
\label{chi}
\chi (\gamma) = 2\, \psi (1 ) - \psi( \gamma) - \psi (1 - \gamma)
\hspace*{1 cm} ,
\end{equation}
with $\psi$ being the Euler $\psi$-function.

The BFKL function (\ref{gareprf}) has poles at $N = 0$ order by order
in perturbation theory. Eq.~(\ref{ktfac}) shows that the poles in the
$\gamma^* \, \gamma^*$ cross section are generated from the ones in
${\cal F}_N$ by integrating the color functions over $k_\perp$.
While these functions are specific to the off-shell photon probe,
the function ${\cal F}$ is universal. The same function contributes to
the small $x$ behavior of the cross sections in hadron-initiated
processes~\cite{hef} via the high energy factorization formulas.

By inserting the representation (\ref{gareprf}) in Eq.~(\ref{ktfac}),
and scaling out the dependence on the photon virtualities, we get
\begin{equation}
\label{gareprsi}
{\sigma}_N ({Q}_A^2, {Q}_B^2) =
{1
\over { 2 \, \pi \, Q_A \, Q_B} } \,
\int_{1 / 2 - i \infty}^{1 / 2 + i \infty} \, {{d \, \gamma} \over
{2 \pi i}} \,
\left( {{ {Q}^2_A } \over { {Q}^2_B} } \right)^{\gamma - 1 / 2 }
\, {1 \over {N - {\bar \alpha_s} \, \chi (\gamma) }}
\, V_1( \gamma) \, V_1 ( 1 - \gamma)
\hspace*{0.8 cm} ,
\end{equation}
where $V_1 ( \gamma)$ is defined as the following
$k_{\perp}$-transform of the photon color function $ G_1$
\begin{equation}
\label{videf}
V_1 ( \gamma) =
\int_0^{\infty} \, {{d \, {{\bf k}}^2} \over { {{\bf k}}^2}} \,
\left( {{ {{\bf k}}^2 } \over { {Q}^2} } \right)^{\gamma - 1} \,
G_1 ( {{ {{\bf k}}^2 } \over { {Q}^2} } ) \hspace*{0.8 cm} .
\end{equation}

The explicit expression of the function $ V_1(\gamma) $ can be
determined by using the representation (\ref{twodimintg}) and
performing the integral transform. The result reads
\begin{equation}
\label{viexp}
V_1 ( \gamma) = \pi
\alpha \, \alpha_s \, \left( \sum_q e^2_q \right) \, {
{ (1 + \gamma) \, ( 2 - \gamma) \,
\Gamma^2 (\gamma) \, \Gamma^2 ( 1 - \gamma) } \over
{(3 - 2 \, \gamma) \, \Gamma ( 3 / 2 + \gamma)
\, \Gamma ( 3 /2 - \gamma) }}
\hspace*{0.8 cm} .
\end{equation}

Eq.~(\ref{gareprsi}), together with the explicit formulas
(\ref{chi}), (\ref{viexp}), gives the leading logarithmic result for
the moments of the $ \gamma^{*} \gamma^{*}$ total cross section. It
sums the $1/N$ poles to the accuracy $( \alpha^2/N ) \times \left(
\alpha_s / N \right)^k $, for any $k$.

The lowest order perturbative contribution, $k = 0$, can be recovered
from the summed formula (\ref{gareprsi}) by expanding the
denominator $\left( N - {\bar \alpha_s} \, \chi (\gamma) \right)^{-1}
$ to the zeroth order in $\alpha_s$. The simple pole $N^{-1}$ is the
Mellin transform of unity, and one can check numerically that the
$\gamma$-integral
\begin{equation}
\label{gareprsi0}
2 \, \pi \, {Q}_A \, {Q}_B \,
{\sigma}^{(0)} ({Q}_A^2, {Q}_B^2) =
\int_{1 / 2 - i \infty}^{1 / 2 + i \infty} \, {{d \, \gamma} \over
{2 \pi i}} \,
\left( {{ {Q}^2_A } \over { {Q}^2_B} } \right)^{\gamma - 1 / 2 }
\, V_1( \gamma) \, V_1 ( 1 - \gamma)
\hspace*{0.8 cm}
\end{equation}
agrees with Eq.~(\ref{sigmarepr}) and Fig.~\ref{s0}.

In general, multiple pole contributions $1 / N^{k+1}$ to the cross
section are obtained by retaining higher orders in the
$\alpha_s$-expansion of Eq.~(\ref{gareprsi}). We see that the general
structure of the coefficients of the leading logarithmic series comes
from both $\chi (\gamma)$ and $V_1 (\gamma)$: the former is a
universal function associated with the BFKL pomeron, while the latter
describes the coupling of the pomeron to a specific physical source.

\subsection{Energy dependence }
\label{subsec:energy}

The total cross section is obtained from Eq.~(\ref{gareprsi}) by
taking the inverse Mellin-Fourier transform (\ref{mellin}). By
evaluating the $N$-integral from the residue at the pole
$N = {\bar \alpha_s} \, \chi (\gamma)$, one gets
\begin{equation}
\label{energy}
{\sigma} (s , {Q}_A^2, {Q}_B^2) =
{1 \over { 2 \, \pi \, Q_A \, Q_B} } \,
\int_{1 / 2 - i \infty}^{1 / 2 + i \infty} \, {{d \, \gamma} \over
{2 \pi i}} \,
\left( {{ {Q}^2_A } \over { {Q}^2_B} } \right)^{\gamma - 1 / 2 }
\left( {s \over {Q^2 }}
\right)^{ {\bar \alpha_s} \, \chi (\gamma) }
\, V_1( \gamma) \, V_1 ( 1 - \gamma)
\hspace*{0.4 cm} .
\end{equation}
Note that this result depends on two mass scales, the scale $\mu^2$ in
$\alpha_s$ and the scale $Q^2$ in the high energy corrections, whose
reliable determination would require a next-to-leading analysis. We
discuss the uncertainties in the leading log result associated with
these scales in Secs.~\ref{sec:scales} and \ref{sec:numerical}.

In the limit $s \to \infty$ the integral (\ref{energy}) is dominated
by the region near $ \gamma = 1 / 2$, where the function $\chi$ has a
saddle point. In the saddle point approximation one obtains
\begin{equation}
\label{saddle}
{\sigma} (s , {Q}_A^2, {Q}_B^2) \simeq
{1 \over { 2 \, \pi \, Q_A \, Q_B} } \,
\, { {{| V_1( 1 / 2) |}^2} \over { \sqrt{2 \, \pi \,
\chi^{\prime \prime} (1 / 2) \, {\bar
\alpha_s} \, \ln \left( s / Q^2 \right) }} }
\,
\left( {s \over {Q^2 }}
\right)^{ {\bar \alpha_s} \, \chi (1 / 2) }
\hspace*{0.8 cm} ,
\end{equation}
with
\begin{equation}
\label{onehalf}
\chi ( 1 / 2 ) = 4 \, \ln 2 \simeq 2.77
\hspace*{0.2 cm} , \hspace*{0.4 cm}
V_1 ( 1 / 2 ) = 9 \, \pi^3
\alpha \, \alpha_s \, \left( \sum_q e^2_q \right)
/ 8 \hspace*{0.2 cm} , \hspace*{0.4 cm}
\chi^{\prime \prime} (1 / 2) =
28 \, \zeta( 3) \simeq 33.66 \hspace*{0.2 cm} .
\end{equation}
Eq.~(\ref{saddle}) shows the asymptotic power behavior characteristic
of the QCD pomeron, $ s^\lambda $ with $\lambda = {\bar \alpha_s} \,
\chi (1 / 2 ) \simeq 2.65\ \alpha_s$. The pre-factor that determines
the normalization of the asymptotic cross section, on the other hand,
depends on the off-shell photon probe, and is controlled by the value
of the function $ V_1 ( \gamma) $ at $\gamma = 1 / 2 $.

If the two photon virtualities are significantly far apart,
corrections of order $\ln (Q^2_A / Q^2_B) / ( \alpha_s \, \ln (s /
Q^2) ) $ need to be taken into account when one calculates the large
$s$ limit. In this case, one finds that the position of the saddle
point is shifted to $\gamma \simeq 1 / 2 - \ln (Q^2_A / Q^2_B) /
\left( {\bar\alpha}_s \, \chi^{\prime \prime} (1 / 2) \, \ln (s /
Q^2) \right)$. Apart from corrections in the pre-factor, the net
effect of evaluating the integral (\ref{energy}) around the shifted
saddle point is to multiply the expression in the right hand side of
Eq.~(\ref{saddle}) by the factor
\begin{equation}
\label{incldiff}
\exp \left[ - { { \ln^2 \left( {Q}^2_A / {Q}^2_B \right) } \over
{ 2 \,
\chi^{\prime \prime} (1 / 2) \, {\bar
\alpha_s} \, \ln \left( s / Q^2 \right) }} \right]
\;\;\; .
\end{equation}
The cross section acquires a gaussian modulation in $ \ln (Q^2_A /
Q^2_B)$ with a width that grows like $\ln s$.

\begin{figure}[htb]
\centerline{ \DESepsf(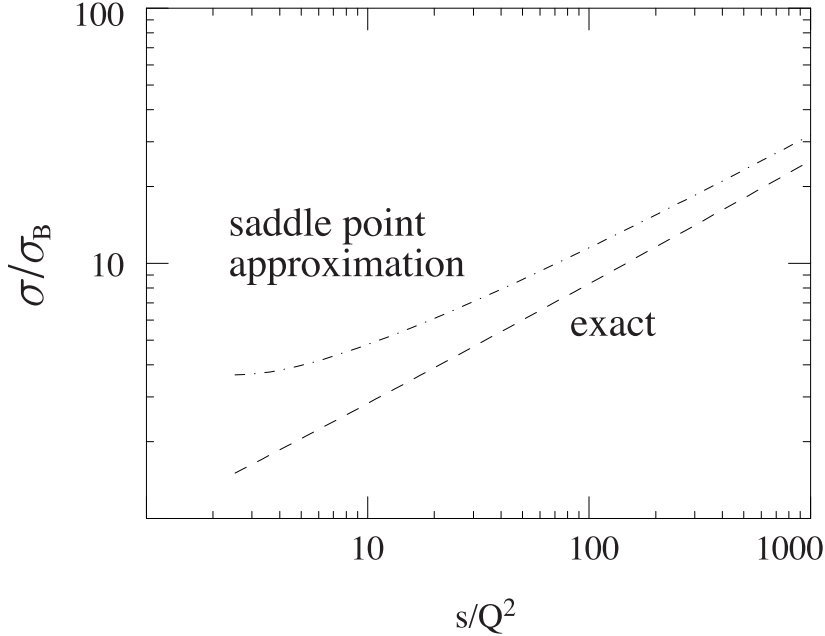 width 13 cm) }
\caption{ The $s/Q^2$ dependence of the $\gamma^* \, \gamma^*$
cross section, Eq.~(\protect\ref{energy}).
We take $Q_A^2 = Q_B^2$, $\alpha_s = 0.2$, and divide
$\sigma$ by the Born cross section, which eliminates the
factor $1/( Q_A Q_B ) $. }
\label{xsect}
\end{figure}

\begin{figure}[htb]
\centerline{ \DESepsf(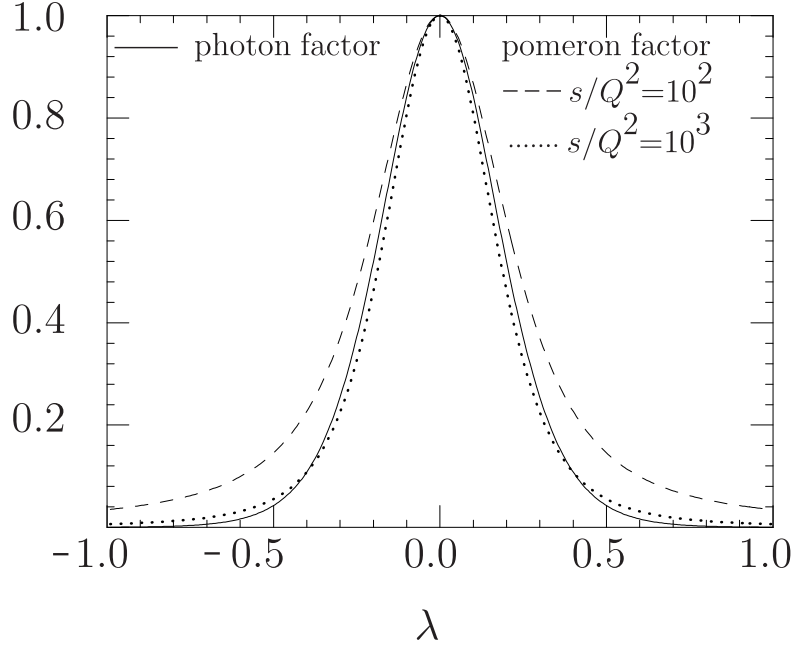 width 13 cm) }
\caption{ The BFKL pomeron factor
$\exp \left[ {\bar \alpha_s} \, \, \left( \chi (\gamma) - \chi(1/2)
\right) \ln (s / Q^2) \right] $ and the photon factor
$V_1(\gamma) \, V_1(1 - \gamma) / V_1^2 (1/2)$ (normalized to the
saddle point) along the contour of integration in
Eq.~(\protect\ref{energy}). We parametrize this contour as $\gamma = 1
/ 2 + i \, \lambda $. We take $\alpha_s = 0.2$, and show the BFKL
pomeron factor for two different values $s/Q^2 = 10^2 $, $s/Q^2 =
10^3 $. }
\label{chi05}
\end{figure}

In the general case, the integral (\ref{energy}) can be performed
numerically. In Fig.~\ref{xsect} we show the result as a function of
$s / Q^2$ for a given choice of the values of the photon virtualities
and the strong coupling. For comparison we also plot the saddle point
formula. As the energy increases the two curves get closer. However,
in the range considered, the sub-asymptotic contributions are still
significant (about $50 \%$ at $Q^2/s
\sim 10^{-1} $, $25 \% $ at $Q^2/s \sim 10^{-3} $). The large size of
the corrections to the saddle point approximation can be mainly
traced back to the fact that the function $V_1( \gamma)$ itself is
rather sharply peaked around $ \gamma = 1 / 2 $. This is illustrated
in Fig.~\ref{chi05}, where we see that, for instance, for
$\alpha_s = 0.2$ and $s / Q^2 = 10^{2}$, the width of the pomeron
factor is still not small compared to the width of the factor
associated with the off-shell photon color function. This effect
accounts for most of the shift in the normalization of the cross
section between the asymptotic and exact evaluation of the leading
logarithmic sum.

\subsection{Summed results for the asymmetry and the longitudinal
cross section}
\label{subsec:asym}

Using the same method described above for the cross section averaged
over the two transverse photon polarizations, one can derive summed
results for each photon polarization $ ( 1 , 2 , L ) $ and for the
polarization asymmetry ${\cal A}$. Denoting by $a$, $a^\prime$ and
$b$, $b^\prime$ the polarization indices for photons $q_A$ and $q_B$,
we write the factorization formula for the moments of the polarized
cross section $\sigma_{ N }^{a a^\prime , b b^\prime }$ as
\begin{equation}
\label{ktfacpol}
\sigma_{ N }^{a a^\prime , b b^\prime } ( Q_A^2, Q_B^2)
=
\int \, {{d^2 \, {{\bf k}}_A} \over { \pi \, {{\bf k}}^2_A}} \,
\int \, {{d^2 \, {{\bf k}}_B} \over { \pi \, {{\bf k}}^2_B}} \,
G^{a a^\prime}({{\bf k}}_A , Q_A) \,
{\widetilde{\cal F}}_N ({{\bf k}}_A, {{\bf k}}_B) \,
G^{b b^\prime}({{\bf k}}_B , Q_B) \,
\;\;\;\;.
\end{equation}
The color functions $G$ have been discussed in
Sec.~\ref{sec:polarization}. The full solution for the Green's
function ${\widetilde{\cal F}}$ of the BFKL pomeron reads \cite{BFKL}
\begin{equation}
\label{gareprfm}
{\widetilde{\cal F}}_N ({{\bf k}}_A, {{\bf k}}_B) =
{ 1 \over { 2 \, \, \pi \, 
\sqrt{ {{\bf k}}^2_A \, {{\bf k}}^2_B} } } \,
\sum_{m = - \infty}^{+ \infty}
\int_{1 / 2 - i \infty}^{1 / 2 + i \infty} \, 
{{d \, \gamma} \over {2 \pi i}}
\, {1 \over {N - {\bar \alpha_s}
\, \chi_m (\gamma ) }}
\,
\left( {{ {{\bf k}}^2_A } 
\over { {{\bf k}}^2_B} } \right)^{\gamma - 1 / 2 }
\, e^{i \, m \, ( \phi_A - \phi_B ) }
\hspace*{0.1 cm}
\end{equation}
with
\begin{equation}
\label{chim}
\chi_m (\gamma ) = 2\, \psi (1 ) - \psi( \gamma + | m | / 2)
- \psi (1 - \gamma + | m | / 2)
\hspace*{1 cm} .
\end{equation}
The function ${\cal F}$ in Eq.~(\ref{gareprf}), relevant to the case
of the unpolarized cross section, is given by the term $ m = 0 $ in
Eq.~(\ref{gareprfm}).

Consider the product $\Sigma = {\overline {\sigma}} \, {\cal A}$ of
the asymmetry times the averaged cross section, with the asymmetry
${\cal A}$ defined in Eq.~(\ref{unplusass}). Due to the azimuthal
integration only the terms with $m = +2 , -2$ in Eq.~(\ref{gareprfm})
contribute. Substituting the color function $G_2$ (see
Eq.~(\ref{gij})) in Eq.~(\ref{ktfacpol}), one finds for the $N$
moments
\begin{equation}
\label{gareprsiasy}
\Sigma_N ({Q}_A^2, {Q}_B^2) =
{1
\over { 16 \, \pi \, Q_A \, Q_B} } \,
\int_{1 / 2 - i \infty}^{1 / 2 + i \infty} \, {{d \, \gamma} \over
{2 \pi i}} \,
\left( {{ {Q}^2_A } \over { {Q}^2_B} } \right)^{\gamma - 1 / 2 }
\, {1 \over {N - {\bar \alpha_s} \, \chi_2 (\gamma ) }}
\, V_2( \gamma) \, V_2 ( 1 - \gamma)
\hspace*{0.8 cm} ,
\end{equation}
where $V_2(\gamma)$ is defined from the
$k_{\perp}$-transform of the function $ G_2$ analogous to
(\ref{videf}), and has the expression
\begin{equation}
\label{viexp2}
V_2 ( \gamma) = 2 \, \pi
\alpha \, \alpha_s \, \left( \sum_q e^2_q \right) \, {
{ \Gamma (1 + \gamma) \, \Gamma ( 2 - \gamma) \,
\Gamma (\gamma) \, \Gamma ( 1 - \gamma) } \over
{(3 - 2 \, \gamma) \, \Gamma ( 3 / 2 + \gamma)
\, \Gamma ( 3 /2 - \gamma) }}
\hspace*{0.8 cm} .
\end{equation}
By inverse Mellin-Fourier transformation, one obtains the summed
formula for the asymmetry in the energy space:
\begin{eqnarray}
\label{energyasy}
{\cal A} (s , {Q}_A^2, {Q}_B^2) &=&
{ 1 \over { {\overline {\sigma}} (s, Q_A^2, Q_B^2) }} \,
{1 \over { 16 \, \pi \, Q_A \, Q_B} }
\nonumber \\
&& \times \,
\int_{1 / 2 - i \infty}^{1 / 2 + i \infty} \, {{d \, \gamma} \over
{2 \pi i}} \,
\left( {{ {Q}^2_A } \over { {Q}^2_B} } \right)^{\gamma - 1 / 2 }
\left( {s \over {Q^2 }}
\right)^{ {\bar \alpha_s} \, \chi_2 (\gamma ) }
\, V_2( \gamma) \, V_2 ( 1 - \gamma)
\hspace*{0.4 cm} ,
\end{eqnarray}
where ${\overline {\sigma}}$ is given by Eq.~(\ref{energy}).

As in the case of the unpolarized cross section, the asymptotic
behavior in the limit $s \to \infty$ is determined by the saddle point
approximation to the integral in Eq.~(\ref{energyasy}). In this case
we observe a negative power law with the energy $s$, controlled by
the value of $\chi_2$ at the saddle point, $\chi_2 (1 / 2 ) = - 4 \,
( 1 - \ln 2 ) \simeq - 1.23 $, indicating that the angular
correlations between the two photons tend to be washed out when BFKL
pomeron exchange dominates.

The contributions to the photon-photon cross section from
longitudinally polarized photons are given by formulas analogous to
Eq.~(\ref{energy}) in terms of different combinations of the
functions $V_1$ and $V_2$, and the same function $\chi$:
\begin{eqnarray}
\label{energylong}
&& {\sigma}^{(LL)} (s , {Q}_A^2, {Q}_B^2) =
{1 \over { 2 \, \pi \, Q_A \, Q_B} } \,
\int_{1 / 2 - i \infty}^{1 / 2 + i \infty} \, {{d \, \gamma} \over
{2 \pi i}} \,
\left( {{ {Q}^2_A } \over { {Q}^2_B} } \right)^{\gamma - 1 / 2 }
\left( {s \over {Q^2 }}
\right)^{ {\bar \alpha_s} \, \chi (\gamma) }
\, V_2( \gamma) \, V_2 ( 1 - \gamma)
\hspace*{0.2 cm} ,
\nonumber \\
&&
{\sigma}^{(TL)} (s , {Q}_A^2, {Q}_B^2) =
{\sigma}^{(LT)} (s , {Q}_A^2, {Q}_B^2)
\nonumber \\
&&
=
{1 \over { 2 \, \pi \, Q_A \, Q_B} } \,
\int_{1 / 2 - i \infty}^{1 / 2 + i \infty} \, {{d \, \gamma} \over
{2 \pi i}} \,
\left( {{ {Q}^2_A } \over { {Q}^2_B} } \right)^{\gamma - 1 / 2 }
\left( {s \over {Q^2 }}
\right)^{ {\bar \alpha_s} \, \chi (\gamma) }
\, V_1( \gamma) \, V_2 ( 1 - \gamma)
\hspace*{0.2 cm} .
\end{eqnarray}
Unlike $V_1 (\gamma) $ in Eq.~(\ref{viexp}), $ V_2$ has simple poles
at $ \gamma = 0 , 1$ (instead of double poles), corresponding to the
non-logarithmic behavior of the color function $G_2$ at $ {{\bf k}}^2
\ll Q^2$, $ {{\bf k}}^2 \gg Q^2$ noted in Sec.~\ref{sec:polarization}.
Replacing functions $V_1$ by functions $V_2$ accounts for the
different size of the longitudinal cross sections with respect to the
purely transverse one. Roughly, one finds
\begin{equation}
\label{magnl}
{\sigma}^{(LL)} \approx 0.05 \; {\sigma}^{(TT)}
\hspace*{0.2 cm} , \hspace*{0.4 cm}
{\sigma}^{(TL)} \approx 0.2 \; {\sigma}^{(TT)}
\hspace*{0.2 cm} .
\end{equation}

\section{Scale dependence and uncertainties of the leading logarithmic
approximation }
\label{sec:scales}

The result (\ref{energy}) for the $\gamma^* \, \gamma^*$ cross section
depends on two mass scales which cannot be determined in a leading
logarithmic analysis: the mass $\mu^2$ at which the running coupling
$\alpha_s$ is evaluated, and the mass $Q^2$ that provides the scale
for the high energy logarithms. The former can be thought of as being
associated with the integrations over the transverse momenta in the
loops contributing to higher order diagrams, while the latter stems
from the longitudinal integrations. A reliable determination of these
scales would require a next-to-leading order calculation. Lacking
this, we provide here qualitative arguments to relate these scales
with the physical hard scales of the problem. In
Sec.~\ref{sec:numerical} we will use these relations to examine
numerically the dependence of the cross section on the scale choices.

A possible choice of the scale $\mu^2$ in the strong coupling is based
on the prescription of Ref.~\cite{BLM}. To apply this prescription,
we consider the lowest order gluon exchange contribution, which we
have discussed in Sec.~\ref{sec:notations}. This is given (see
Eq.~(\ref{convsigma})) by the integral in $d {{\bf k}}^2$ of two $G$
factors, each of which is proportional to $\alpha_s$, $ G = \alpha_s(
\mu^2) \, {\cal G} $. We first compute the quark loop contribution to
the gluon propagator, renormalized in the ${\overline {\mbox {MS}}}$
scheme:
\begin{equation}
\label{pai}
\Pi ({\bf k}^2) = {\tilde \beta}_0 \,
\left[\ln ( {{\bf k}}^2 / \mu^2 ) + C\right] \;\;\;\;,
\end{equation}
where ${\tilde \beta}_0 = - 1 / (6 \, \pi)$ is the contribution to
$ \beta_0 = (33 - 2 \, N_f ) / (12 \, \pi)$ from one quark loop, and
$C = - 5 /3$. Inserting a quark loop into the gluon propagator in the
lowest order diagram amounts to replacing $\alpha_s (\mu^2)$ by
\begin{equation}
\label{alphas}
\alpha_s (\mu^2) \left[ 1 - {\tilde \beta}_0 \,
\alpha_s (\mu^2) \,
\ln ( {{\bf k}}^2 \, e^{ C }/ \mu^2 ) \right]
\end{equation}
(as in \cite{BLM} this can be regarded as a contribution to an
effective coupling $ \alpha_{eff} ({{\bf k}}^2)$). Now, following
\cite{BLM} we choose the scale $\mu^2$ so that the quark loop
contribution vanishes after integrating over ${{\bf k}}^2$:
\begin{equation}
\label{vanish}
\int_0^{\infty} { {d {{\bf k}}^2} \over { ( {{\bf k}}^2 )^2 }} \,
{\cal G}({{\bf k}}^2/ Q^2_A) \,
{\cal G}({{\bf k}}^2/ Q^2_B) \,
\ln ( {{\bf k}}^2 \, e^{ C }/ \mu^2 )
= 0 \;\;\;\;.
\end{equation}
This is the same procedure that applies in the case of the abelian
theory. Using the representation (\ref{twodimintg}) for $G$, we are
led to calculate an integral of the form
\begin{equation}
\label{meanlog}
\int_0^{\infty} {d {{\bf k}}^2}
{1 \over {
\lambda_A \, (1 - \lambda_A) \,{{\bf k}}^2 + z_A \, (1-z_A) \, Q^2_A
} } \,
{1 \over {
\lambda_B \, (1 - \lambda_B) \,{{\bf k}}^2 + z_B \, (1-z_B) \, Q^2_B
} } \,
\ln \left( {{ {{\bf k}}^2 \, e^{ C } } \over {\mu^2}} \right)
\;\;\;\;.
\end{equation}
Exploiting the symmetry of the integrand under the transformation
${{\bf k}}^2/ Q_A \, Q_B \to Q_A \, Q_B / {{\bf k}}^2$ and
interchange of the variables $z_A \leftrightarrow \lambda_B$, $z_B
\leftrightarrow \lambda_A$, one can show that the condition
(\ref{vanish}) is satisfied by
\begin{equation}
\label{cmu}
\mu^2 = c_\mu \,
Q_A \, Q_B \hspace*{0.5 cm} , \hspace*{0.6 cm}
c_\mu = e^{-5/3} \hspace*{0.5 cm} .
\end{equation}
Note that, when the two photon virtualities $Q_A^2$ and $Q_B^2$ are
far apart from each other, the prescription \cite{BLM} picks out a
scale for $\alpha_s$ which is neither of the order of the big
virtuality nor of the order of the small one, but rather is
proportional to the geometric mean $\sqrt{ Q_A^2 \, Q_B^2}$. The value
of the proportionality coefficient $c_\mu$ in Eq.~(\ref{cmu}) is
specific to the subtraction scheme (${\overline {\mbox {MS}}}$)
chosen to define the quark loop insertion.

The argument given above applies to the factors of $\alpha_s$ that
appear in the Born approximation to the high energy cross section.
The result (\ref{energy}), however, also contains a dependence on the
running coupling through the higher order factor $s^{ \alpha_s \,
\chi}$ associated with the solution of the BFKL equation. For the
scale in $\alpha_s$ in this case one does not have such a simple
argument as the one described above. For the numerical estimates in
Sec.~\ref{sec:numerical} we will make the assumption that the same
value of $\mu^2$ also controls the running coupling in the BFKL
factor. In Sec.~\ref{sec:numerical} we will check the numerical
effect of varying this scale.

We now consider the mass $Q^2$ that provides the scale for the large
energy logarithms $\left( \alpha_s \ln (s / Q^2) \right)^k$ (see
Eq.~(\ref{expansion})). To estimate this scale, we observe that the
rapidity of gluons exchanged in the rungs of the BFKL ladders should
lie between the rapidity $y_A$ of the quark $p_A$ (produced by the
photon $q_A$) and the rapidity $y_B$ of the quark $p_B$ (produced by
the photon $q_B$). This gives rise to integrations over the rapidity
intervals
\begin{equation}
\label{rapint}
\int_{y_B}^{y_A} dy = y_A - y_B \hspace*{0.3 cm} .
\end{equation}
The logarithms of the energy $s$ are generated precisely by these
integrals. Estimating the size of the rapidity intervals thus allows
us to estimate the scale $Q^2$.

Expressing the ``$+$'' and ``$-$'' momentum components of the quark
$p_A$ as
\begin{equation}
\label{aplusminus}
p_A^{+} = z_A \, q^{+}_A \hspace*{0.3 cm} , \hspace*{0.5 cm}
p_A^{-} = {{{\bf p}}_{\perp \, A}^2 \over {2 \; p^{+}_A }} 
\hspace*{0.5 cm} ,
\end{equation}
we write its rapidity as
\begin{equation}
\label{ya}
y_A = {1 \over 2} \ln \left( {{ p_A^{+} }\over { p_A^{-} }}
\right) = {1 \over 2} \ln \left( {{ 2 \, z_A^2 \, ( q^{+}_A )^2 } 
\over { {{\bf p}}_{\perp \, A}^2 }} \right) \hspace*{0.5 cm} .
\end{equation}
Similarly, for the quark $p_B$ we have
\begin{equation}
\label{yb}
y_B = {1 \over 2} \ln \left( {{ p_B^{+} }\over { p_B^{-} }}
\right) = -
{1 \over 2} \ln \left( {{ 2 \, z_B^2 \, ( q^{-}_B )^2 } \over
{ {{\bf p}}_{\perp \, B}^2 }} \right) \hspace*{0.5 cm} .
\end{equation}
Taking the difference between these rapidities gives
\begin{equation}
\label{rapid}
y_A - y_B = \ln \left( { { z_A \, z_B \, s} \over
{ |{{\bf p}}_{\perp \, A}| \, |{{\bf p}}_{\perp \, B}| } }
\right) \;\;\;\;\;.
\end{equation}
The average transverse momenta ${{\bf p}}_{\perp \, A}$, ${{\bf
p}}_{\perp\, B}$ carried by the quarks are of the order of the photon
virtualities $Q_A$, $Q_B$. For the longitudinal momentum fractions
$z_A$, $z_B$, we assume a typical maximum value of the order
$z_{\rm max} \sim 0.1 $ in the high energy region. Using these
estimates, we obtain
\begin{equation}
\label{y100}
y_A - y_B = \ln \left( { { s \, z_{\rm max}^2 } \over
{ Q_A \, Q_B } }
\right) \;\;\;\;\;.
\end{equation}
We thus identify the scale
\begin{equation}
\label{cqu}
Q^2 = c_Q \, Q_A \, Q_B \hspace*{0.5 cm} , \hspace*{0.6 cm}
c_Q = 1 / z_{\rm max}^2 = 10^2 \hspace*{0.5 cm} .
\end{equation}

\section{Limitations on the BFKL pomeron approach at very large
energy}
\label{sec:limitations}

The transverse-momentum integrations in the factorization formula
(\ref{ktfac}) are dominated by values of ${{\bf k}}^2_A , {{\bf
k}}^2_B$ of the order of the photon virtualities. (For the purpose of
this section we will take the photon virtualities to be equal, ${Q}^2_A
= {Q}^2_B$.) As a result, for sufficiently off shell photons the
dominant contribution to the cross section comes from short distances,
and the evaluation of Eq.~(\ref{ktfac}) gives rise to a finite result
in perturbation theory.  However, there are limitations on the
perturbative treatment that are intrinsic to the BFKL equation.  These
limitations come from the region of very high $s$. Although an accurate
understanding of these effects is an open problem~\cite{bjfutu} that
goes beyond the scope of this work, one can nevertheless make some
rough estimates. We discuss them in this section. In order to keep the
notation simple, we do not distinguish here between the scales $Q_A
Q_B$, $Q^2 = c_Q Q_A Q_B$, and $\mu^2 = c_\mu Q_A Q_B$, calling all of
these simply $Q^2$.

It is known from the structure of the BFKL equation that, even if the
incoming ${{\bf k}}^2_A$, ${{\bf k}}^2_B$ are large, say, ${{\bf
k}}^2_A , {{\bf k}}^2_B \sim  Q^2 $, the typical
transverse momenta in the gluon ladders contributing to the function
${\cal F}_N ({{\bf k}}_A, {{\bf k}}_B)$ (see Eq.~(\ref{gareprf})) may
diffuse away from $Q^2$ as the energy becomes very large. The
diffusion coefficient in $\ln {{\bf k}}^2$ can be read directly from
the asymptotic solution to the BFKL equation (or, equivalently, from
the exponential term (\ref{incldiff}) in the $\gamma^* \, \gamma^*$
cross section) and is given by
\begin{equation}
\label{diffu}
\rho_{\rm diff} = 2 \,
\chi^{\prime \prime} (1 / 2) \, {\bar
\alpha_s} \, \ln \left( s / Q^2 \right) \,\;\; .
\end{equation}
As a result, the distribution of the transverse momenta in the BFKL
ladders is a gaussian in $\ln {{\bf k}}^2$ centered around $Q^2$ with
a width proportional to $\rho_{\rm diff}$~\cite{bardiff}. With
increasing $s$ the distribution broadens, and one becomes sensitive
to the region of small transverse momenta.

A self-consistency check of the perturbative treatment requires that
the contribution from transverse momenta of the order of
$\Lambda_{QCD}$ be suppressed. In order to keep away from momenta of
this order one has to have
\begin{equation}
\label{difflimit}
\ln^2 ( Q^2 / \Lambda^2 ) / \left(
2 \,
\chi^{\prime \prime} (1 / 2) \, {\bar
\alpha_s} \, \ln ( s / Q^2 ) \right)
\gtrsim 1 \;\;\;\; .
\end{equation}
Identifying $\alpha_s$ with the strong coupling evaluated at the
scale $Q^2$, $\alpha_s \simeq \left( \beta_0 \,
\ln (Q^2 / \Lambda^2 ) \right)^{-1}$, one gets
\begin{equation}
\label{difflimit1}
\alpha_s (Q^2) \, \ln ( s / Q^2 ) \lesssim c_1 / \alpha_s^2 (Q^2)
\;\; , \hspace*{1.2 cm}
c_1 = \pi / \left( 2 \,
\chi^{\prime \prime} (1 / 2) \, \beta_0^2 \, C_A \right)
\approx 1 / 30
\;\, .
\end{equation}
For any given $Q^2$, this can be read as an upper bound on the domain
of energies $s$ in which we expect the perturbative approach based on
the BFKL pomeron to be reliable. Observe that, for small values of
$\alpha_s$, this is not so stringent a constraint. However, because
the BFKL function $\chi$ has a large second derivative at the saddle
point, the numerical value of the coefficient $c_1$ is small.
Therefore, the limit (\ref{difflimit1}) may become relevant unless
$Q^2$ is very large. 

The BFKL equation is also known to give rise to violation of the
unitarity bound at asymptotically large energies. The growth of the
cross section predicted by the BFKL equation cannot continue
indefinitely, and unitarity corrections must arise to slow it down.
Roughly, these effects are expected to become important when the
calculated cross section is bigger than the naive geometrical cross
section $ 1 / Q^2$. A careful discussion may be found in
Ref.~\cite{muni}. The simplest estimate, $\sigma \lesssim 1/Q^2$,
yields, using Eq.~(\ref{saddle}),
\begin{equation}
\label{unitlimit}
{ {\alpha_s^2} \over { \sqrt{2 \, \pi \,
\chi^{\prime \prime} (1 / 2) \, {\bar
\alpha_s} \, \ln \left( s / (Q^2 ) \right) }} }
\,
\left( {s \over {Q^2 }}
\right)^{ {\bar \alpha_s} \, \chi (1 / 2) }
\lesssim 1 \;\;\;\; ,
\end{equation}
where we have neglected factors from the term in $V_1 ( 1 / 2)$ in
Eq.~(\ref{saddle}). Assuming, as we did before,
$\alpha_s$ to be evaluated at the scale $Q^2$, we can write
\begin{equation}
\label{unitlimit1}
\alpha_s (Q^2) \, \ln ( s / Q^2 )
\lesssim c_2 \,
\ln (1 / \alpha_s^2 (Q^2) ) + A
\;\; , \hspace*{1.2 cm}
c_2 = \pi / \left( C_A \, \chi (1 / 2) \right)
\approx 1 / 3
\;\, .
\end{equation}
In the term $A$ we collect contributions arising from the factor in
the square root in Eq.~(\ref{unitlimit}) as well as from the factor
$V_1$ in Eq.~(\ref{saddle}).

The different functional dependence on $\alpha_s$ in the right hand
sides of the inequalities (\ref{difflimit1}) and (\ref{unitlimit1})
implies that, for small enough values of $\alpha_s$, the unitarity
limit (\ref{unitlimit1}) is more stringent than the diffusion limit
(\ref{difflimit1}). This suggests that for sufficiently high $Q^2$ it
should be possible to study unitarization in a purely perturbative
context~\cite{muni}.

On the other hand, the coefficients $c_1$ and $c_2$ are significantly
different in size. This may make the two bounds (\ref{difflimit1})
and (\ref{unitlimit1}) rather comparable for moderate values of $Q^2$.
In addition, inspection of Eqs.~(\ref{saddle}), (\ref{unitlimit})
suggests that the term $A$ in Eq.~(\ref{unitlimit1}) is not
necessarily negligible, and may contribute to push the onset of
unitarity corrections further away.

As to the impact on experimental studies at future $e^+ e^-$
colliders, we observe that unitarity corrections should set in when
the cross section has grown to be much larger than the Born value
$\sim \alpha_s^2 / Q^2$. At a future $e^+ e^-$ collider one may thus
expect to see the rise of the cross section with $s / Q^2$. Possibly,
one may see this rise slow as unitarity corrections become
important~\cite{bdl}.

\section{Virtuality dependence and relationship with deeply inelastic
scattering }
\label{sec:virtuality}

If we let $Q_A^2$ be much larger than $Q_B^2$ in Eq.~(\ref{energy}),
we obtain a result that describes small $x$ deeply inelastic
scattering from a transversely polarized photon~\cite{sas} whose
virtuality $Q_B^2$ is sufficiently large to allow the use of
perturbation theory to analyze its decomposition into quarks. The
structure function $F_1(x,Q^2)$ (which is the same as $F_T$) is
related to the virtual photon-photon cross section by
\begin{equation}
F_1(x,Q_A^2) = {1 \over (2 \pi)^2 \alpha}\ Q_A^2\ {\sigma} (s ,
{Q}_A^2, {Q}_B^2)
\end{equation}
with
\begin{equation}
s = {Q_A^2 \over x} \;\;\; .
\end{equation}
Thus Eq.~(\ref{energy}) gives a leading logarithmic summation for
$F_1$ at small $x$:
\begin{equation}
\label{F1}
F_1(x,Q_A^2) =
{1 \over (2 \pi)^3 \alpha } \,
\int_{1 / 2 - i \infty}^{1 / 2 + i \infty} \, {{d \, \gamma} \over
{2 \pi i}} \,
\left( {{ {Q}^2_A } \over { {Q}^2_B} } \right)^{\gamma}
\left( Q_A^2 \over x\,Q^2 \right)^{ {\bar \alpha_s} \,
\chi (\gamma) }
\, V_1( \gamma) \, V_1 ( 1 - \gamma)
\hspace*{0.4 cm} .
\end{equation}
The scale $Q^2$ is not fixed by a leading logarithmic calculation. A
sensible choice for $Q^2$ based on a qualitative argument has been
discussed in Sec.~\ref{sec:scales}. For the purpose of the present
section we simply leave it undetermined.

Consider the perturbative expansion of Eq.~(\ref{F1}),
\begin{equation}
\label{F1pert}
F_1(x,Q_A^2) =
{1 \over (2 \pi)^3 \alpha } \,
\int_{1 / 2 - i \infty}^{1 / 2 + i \infty} \, {{d \, \gamma} \over
{2 \pi i}} \,
\left( {{ {Q}^2_A } \over { {Q}^2_B} } \right)^{\gamma}
\left[ 1 + {\bar \alpha_s} \, \chi (\gamma)
\ln\left(Q_A^2 \over x\, Q^2 \right) + \cdots \right]
\, V_1( \gamma) \, V_1 ( 1 - \gamma)
\hspace*{0.1 cm} .
\end{equation}
The functions $V_1( \gamma)$, $V_1 ( 1 - \gamma)$ and $\chi (\gamma)$
have poles at integer values of $\gamma$. Thus this expression
contains contributions proportional to $(Q_A^2)^0$ times logarithms,
$(Q_A^2)^{-1}$ times logarithms, and so forth. We extract the leading
twist contribution, the part proportional to $(Q_A^2)^0$ times
logarithms, by rewriting the integral as an integral over a contour
$\cal C$ that encircles the singularity at $\gamma = 0$ plus an
integral over a contour from $-1/2 - i\infty$ to $-1/2 + i\infty$.
The integral over the contour $\cal C$ is the leading twist
contribution. We discard the integration over the contour from $-1/2
- i\infty$ to $-1/2 + i\infty$, which contains the higher twist
contributions. Thus the leading twist contribution to $F_1$ is
\begin{eqnarray}
\label{F1pertLT}
F_1^{\rm L.T.}(x,Q_A^2) &=&
{1 \over (2 \pi)^3 \alpha } \,
\int_{\cal C} \, {{d \, \gamma} \over
{2 \pi i}} \,
\left( {{ {Q}^2_A } \over { {Q}^2_B} } \right)^{\gamma}
\left[ 1 + {\bar \alpha_s} \, \chi (\gamma)
\ln\left(Q_A^2 \over x\, Q^2 \right) + \cdots \right]
\, V_1( \gamma) \, V_1 ( 1 - \gamma)
\nonumber\\
&=&
{1 \over (2 \pi)^3 \alpha }\
{\lower 6 pt \hbox{\vbox{\hbox{$\rm Res$}
\kern -6 pt \hbox{$\scriptstyle \,\gamma = 0$}}}}
\,
\left( {{ {Q}^2_A } \over { {Q}^2_B} } \right)^{\gamma}
\left[ 1 + {\bar \alpha_s} \, \chi (\gamma)
\ln\left(Q_A^2 \over x\, Q^2 \right) + \cdots \right]
\, V_1( \gamma) \, V_1 ( 1 - \gamma)
\hspace*{0.1 cm} .
\end{eqnarray}

It is straightforward to determine the perturbative coefficients. The
function $V_1(\gamma)$ has a double pole at $\gamma = 0$, as does the
function $V_1(1-\gamma)$. Thus there are three powers of $\ln(Q_A^2 /
Q_B^2)$ at leading order in $\alpha_s$, which is $\alpha_s^2$ since
$V_1 \propto \alpha_s$. We find
\begin{eqnarray}
\label{triplelog}
F_1^{\rm L.T.}(x,Q_A^2) &=&
{{4\, \alpha \, \alpha_s^2 } \over {9 \, \pi^3 }}
\, \left( \sum_q e^2_q \right)^2
\, \left[ {1 \over 3} \,
\ln^3\! \left( {{Q_A^2} \over {Q_B^2}} \right) +
{7 \over 3} \,
\ln^2\! \left( {{Q_A^2} \over {Q_B^2}} \right)
+
\left( {119 \over 6} - 4 \, \zeta ( 2 ) \right) \,
\ln\! \left( {{Q_A^2} \over {Q_B^2}} \right)
\right.
\nonumber \\
&& \left.
+ \left( {1063 \over 27} - {28 \over 3} \, \zeta ( 2 ) \right) \,
+ {\cal O} \left( \alpha_s \right) \right]
\hspace*{0.2 cm} , \hspace*{0.4 cm} \zeta ( 2 ) \simeq 1.64
\hspace*{0.2 cm}.
\end{eqnarray}

If we restore the summation of the leading logs of $1/x$, we obtain
\begin{equation}
\label{F1LT}
F_1^{\rm L.T.}(x,Q_A^2) =
{1 \over (2 \pi)^3 \alpha } \,
\int_{\cal C} \, {{d \, \gamma} \over
{2 \pi i}} \,
\left( {{ {Q}^2_A } \over { {Q}^2_B} } \right)^{\gamma}
\left( {Q_A^2 \over x{Q^2 }}
\right)^{ {\bar \alpha_s} \, \chi (\gamma) }
\, V_1( \gamma) \, V_1 ( 1 - \gamma)
\hspace*{0.4 cm} .
\end{equation}
This is not a simple function. However, one can easily perform the
integration numerically.

\begin{figure}[htb]
\centerline{ \DESepsf(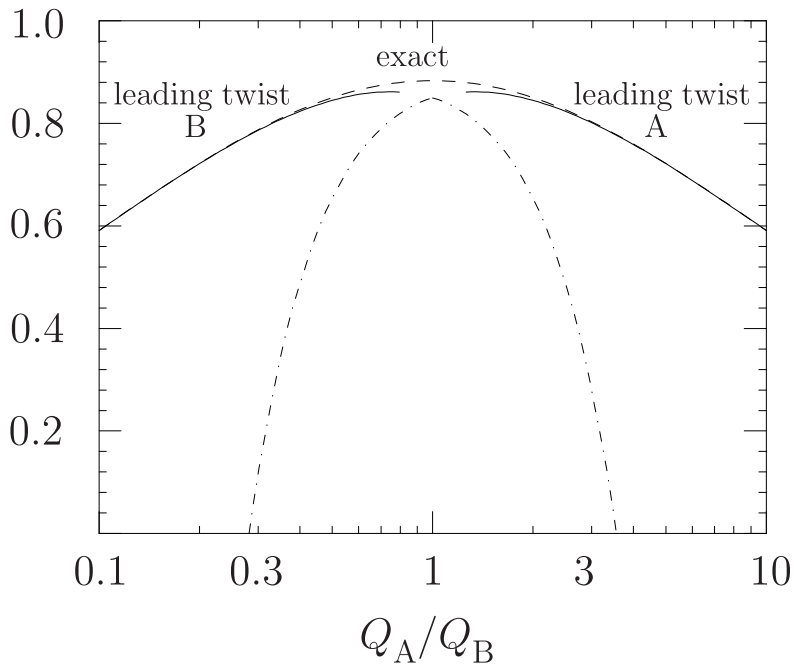 width 13 cm) }
\caption{The full and the leading-twist approximation to the
virtual photon cross section. We plot the cross section divided
by $16 \alpha^2 \alpha_s^2 (\sum_q e_q^2)^2/ (Q_AQ_B)$, as in
Fig.~\protect\ref{s0}.}
\label{twist}
\end{figure}

One interesting result is that we observe what used to be called
``precocious Bjorken scaling''. That is, the leading twist
approximation begins to be accurate at values of $Q_A^2$ that are not
really very large compared to the only other scale in the problem,
$Q_B^2$. We illustrate this in Fig.~\ref{twist}, working at order
$\alpha_s^2$. We compare the leading twist approximation
that applies for $Q_A^2 \gg Q_B^2$, the leading twist approximation
that applies for $Q_B^2 \gg Q_A^2$, and the full result. First we
define $x_A = Q_A^2/s$ and plot
\begin{equation}
\sigma_{\rm L.T.-A}(s,Q_A^2,Q_B^2) =
{{(2 \pi)^2 \alpha}\over Q_A^2}\
F_1^{\rm L.T.}(x_A,Q_A^2)
\end{equation}
versus $ Q_A/Q_B$. Then we define $x_B = Q_B^2/s$ and plot
\begin{equation}
\sigma_{\rm L.T.-B}(s,Q_A^2,Q_B^2) =
{{(2 \pi)^2 \alpha}\over Q_B^2}\
F_1^{\rm L.T.}(x_B,Q_B^2)
\end{equation}
Finally, we plot $\sigma(s,Q_A^2,Q_B^2)$ at the leading order in
$\alpha_s$ without approximation. We see that the leading twist
approximation for $Q_A^2 \gg Q_B^2$ is quite good down to $Q_A^2 =
Q_B^2$, while the leading twist approximation that applies for $Q_B^2
\gg Q_A^2$ is quite good down to $Q_B^2 = Q_A^2$. In fact, at $Q_A^2
= Q_B^2$, both approximations are quite good. A similar behavior is
observed if we include higher orders in $\alpha_s$.

\section{Regge factorization}
\label{sec:regge}

The QCD result for $\gamma^* \, \gamma^* $ scattering can be compared
with expectations for the structure of the high energy cross section
based on traditional Regge theory~\cite{regge}. In Regge theory, to
analyze the elastic scattering of particles $A$ and $B$ one considers
the singularity structure of the amplitude $ A (s, t)$ in the complex
angular momentum plane. The simplest case is a pole in the angular
momentum $j$ plane located at a position dependent on $t$,
$ 1 / \left( j - \alpha (t) \right) $. One then obtains the
asymptotic behavior $s^{\alpha(t)}$ for $s \to \infty$ and $t$ fixed,
and the amplitude takes the factorized form
\begin{equation}
\label{reggeampl}
A ( s , t )
\sim \beta_{A } (t) \,
s^{\alpha (t) } \, \beta_{B } (t)
\hspace*{0.4 cm} .
\end{equation}
Here $\beta_{A}$, $\beta_{B}$ are functions of the transferred
momentum, associated respectively with the couplings of particles $A$
and $B$ to the reggeon whose trajectory is ${\alpha (t)}$.

For the case of the $\gamma^* \, \gamma^*$ total cross section,
this would correspond to the structure
\begin{equation}
\label{reggebehav}
\sigma_{\gamma^* \, \gamma^*} ( s , Q^2_A , Q^2_B )
\sim \beta_{A } ( t=0 ; Q^2_A) \,
s^{\alpha ( 0 ) - 1 } \, \beta_{B } ( t=0 ; Q^2_B)
\hspace*{0.4 cm} ,
\end{equation}
where we have used the optical theorem to relate the total cross
section to the imaginary part of the forward elastic amplitude, and we
have let $\beta_A$ and $\beta_B$ depend on the photon virtualities
in the case of off-shell photons.

However, it has been long known, from various phenomenological and
theoretical considerations~\cite{regge}, that this structure cannot
be exactly true, and strong-interaction scattering at high energy has
to have a more complicated singularity structure than a pole, such as
moving or fixed cuts. In this case, one does not expect the
factorized form for the cross section to hold.

If we now turn to the QCD result (see Eq.~(\ref{gareprsi})), we may
ask what kind of singularities the amplitude has in the angular
momentum plane. The BFKL pomeron is known to give rise to a (fixed)
branch point singularity~\cite{BFKL}. To see this, let us consider the
moments of the $\gamma^* \, \gamma^* $ cross section. At leading
level, we are allowed to identify the moment $N$ with the complex
angular momentum $j$~\cite{regge}. Eq.~(\ref{gareprsi}) is written in
terms of an integral in the complex $\gamma$-plane. The integrand has
the pole $1 / \left( N - {\bar \alpha}_s \, \chi (\gamma) \right) $
and one factor of $V$ for the coupling of the gluon system to the
quarks in each photon. Thus the integrand has a factorized structure.
To understand the angular-momentum singularity structure, one should
see what becomes of the pole $N = {\bar \alpha}_s \, \chi (\gamma)$
after $\gamma$-integration.

For $Q_A^2 > Q_B^2$, the $\gamma$-integral is well approximated by
the residue at the rightmost pole ${\bar \gamma}$ to the left of the
integration contour. As we have seen in the previous section,
numerically this approximation turns out to be fairly good down to
values of $Q_A^2 / Q_B^2$ just above $1$. In this approximation one
gets
\begin{equation}
\label{fnlead}
{\sigma}_N (Q_A^2, Q_B^2) \sim
{ 1 \over { 2 \, \pi \, Q_A^2 } } \,
C_N \,
\exp \left[ {\bar \gamma}_N \, \ln (Q_A^2 / Q_B^2)
\right]
\;\;, \;\;\; C_N = - {{V_1 ({\bar \gamma}_N ) \, 
V_1 (1- {\bar \gamma}_N )}
\over {
{\bar \alpha}_s \, \chi^{\prime} ({\bar \gamma}_N ) } }
\;\;\;.
\end{equation}
The leading pole ${\bar \gamma}_N$ is determined by the equation $N -
{\bar \alpha}_s \, \chi ({\bar \gamma}) = 0$. At $N = {\bar
\alpha}_s\, \chi ( 1 / 2) $ it has a square-root branch point
singularity:
\begin{equation}
\label{sqroot}
{\bar \gamma}_N \sim {1 \over 2} -
\sqrt{ { {2 \, \left( N - N_0 \right)} \over { {\bar \alpha}_s
\, \chi^{\prime \prime} ( 1 / 2) } } } \;\;, \;\;\;
N \to N_0 \equiv {\bar \alpha}_s
\, \chi ( 1 / 2)
\;\;\;.
\end{equation}
Correspondingly, the residue $C_N$ is also singular:
\begin{equation}
\label{singres}
C_N \sim |V_1 (1 / 2)|^2 / \sqrt{ 2 \, {\bar \alpha}_s
\, \chi^{\prime \prime} ( 1 / 2) \,
\left( N - N_0 \right) }
\;\;, \;\;\;
N \to N_0
\;\;\;.
\end{equation}

Therefore, the QCD result implies a branch point rather than a simple
pole in the angular momentum plane. As a consequence, we do not
obtain a Regge-factorized form for $\sigma ( s , Q_A^2 , Q_B^2 ) $.

It is of interest, however, to see by how much this factorization is
violated. We first consider the asymptotic formula (\ref{saddle}),
obtained by using the saddle point approximation around
$\gamma = 1 / 2$. In this case, provided the scale $Q^2$ is fixed by
$ Q^2 = c_Q \, Q_A \, Q_B $, as discussed in Sec.~\ref{sec:scales},
and the QCD coupling $\alpha_s$ is fixed, an approximate form of
Regge factorization is recovered. The piece which violates this
factorization in Eq.~(\ref{saddle}) is proportional to the square
root of a logarithm, and is therefore a slowly varying function. A
more substantial source of factorization breaking comes in when one
takes into account the correction due to the position of the saddle
point drifting away from $1 / 2$, see Eq.(\ref{incldiff}).

For the exact leading log result (\ref{energy}), a possible way to
quantify the deviation from the Regge-factorized behavior is to look
at the ratio
\begin{equation}
\label{reggeratio}
R = { { \left[ \sigma_{\gamma^* \, \gamma^*} ( s , Q^2_A , Q^2_B )
\right]^2 } \over { \sigma_{\gamma^* \, \gamma^*} ( s , Q^2_A , Q^2_A )
\, \sigma_{\gamma^* \, \gamma^*} ( s , Q^2_B , Q^2_B ) } }
\hspace*{0.5 cm} .
\end{equation}
If Regge factorization holds, this quantity should be equal to $1$.
On the other hand, from Eq.~(\ref{energy}) we see that
$R$ goes like $R \sim Q_B^2 / Q_A^2$ for $Q_A^2 \gg Q_B^2$, and $R
\sim Q_A^2 / Q_B^2$ for $Q_B^2 \gg Q_A^2$. That is, for $Q_A$ and
$Q_B$ sufficiently far apart the Regge-factorized form breaks down.

\begin{figure}[htb]
\centerline{ \DESepsf(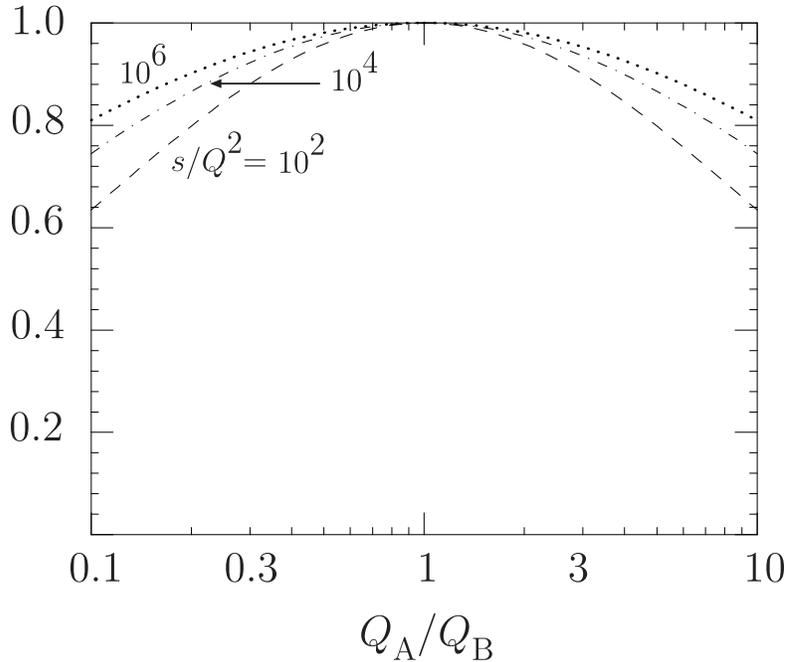 width 13 cm) }
\caption{The $Q_A/Q_B$ dependence of the ratio $R$ defined in
Eq.~(\protect\ref{reggeratio}), at different values of $s / Q^2$. }
\label{fratio}
\end{figure}

In Fig.~\ref{fratio}, we plot $R$ as a function of the ratio of the
photon virtualities $Q_A/Q_B$ for different values of the energy. It
is interesting to observe that, for typical parameter values, $R$
varies by not more than 40\% when $Q_A/Q_B$ varies from 0.1 to 10. We
take this as an indication that an approximate Regge factorization
holds numerically for the exact integral (\ref{energy}) if $Q_A/Q_B$
is not too large or too small.

\section{Soft scattering and hard scattering}
\label{sec:softandhard}

The calculation of the high energy photon-photon cross section
discussed so far is a perturbative calculation, based on the
dominance of short distances for large photon virtualities. When the
photon virtualities decrease, one goes out of the region of validity
of the perturbative approach. As the photons go near the mass shell,
the high energy scattering process is expected to become dominated by
soft interactions. In this regime one is not able to calculate in
QCD, and in order to have a (phenomenological) description of the
cross section, one rather has to rely on models for
strong-interaction scattering based on Regge theory.

For on-shell photons, the Regge factorization hypothesis allows us
to relate the photon-photon total cross section to the photon-proton
and proton-proton cross sections, as follows
\begin{equation}
\label{ppgap}
\sigma_{ {\gamma} {\gamma} } \approx
{ \sigma_{ {\gamma} {p} } \; \sigma_{ {\gamma} {p} } }
/ {\sigma_{ p \, p }}
\hspace*{0.8 cm} .
\end{equation}
Assuming the values $\sigma_{ {\gamma} \, {p} }\approx 0.1 \; {\mbox
{mb}}$, $\sigma_{ p \, p } \approx 40 \; {\mbox {mb}}$, one gets
$\sigma_{ {\gamma} {\gamma} } \approx 250 \; {\mbox {nb}} $. For
virtual photons with small $Q_A$ and $Q_B$, the fall-off of the cross
section can be estimated from vector meson dominance:
\begin{equation}
\label{vmd}
\sigma_{ {\gamma^{*}} {\gamma^{*}} } \sim \left( {{M^2_\rho} \over
{{M^2_\rho} + Q_A^2}} \right)^2 \;\,
\left( {{M^2_\rho} \over
{{M^2_\rho} + Q_B^2}} \right)^2 \;\,
\sigma_{ {\gamma} {\gamma} } \hspace*{0.8 cm} .
\end{equation}
As the photon virtualities increase, the cross section, instead of
continuing to fall like $1 / Q^8$, should begin to fall more slowly.
At large photon virtualities (of the order of a few ${\mbox { GeV}}$,
or bigger), it should go over to the perturbative scaling behavior in
Eq.~(\ref{energy}), $\sigma \propto 1 / Q^2$ at fixed $s / Q^2$.

Note that the $1 / Q^2$ behavior could not be obtained in the
framework of the Regge factorization (\ref{ppgap}) even if one
assumed perturbative scaling in each one of the photon cross section
factors, that is, even if one assumed $\sigma_{\gamma p} \propto 1/
Q^2$. This is the counterpart (at the level of hadronic cross
sections) of the effect of the deviation from unity observed for the
ratio $R$ in the previous section (see Eq.~(\ref{reggeratio})). In
fact, experimental data on the ${\gamma \, p}$ cross section for
large photon virtuality are now available in the region of high
energies from the measurements of small-$x$ deeply inelastic
scattering at HERA. The above observation amounts to saying that,
even if one used the data for $\sigma_{\gamma p}$, the relation
(\ref{ppgap}) would not lead to the correct perturbative QCD result
for the $\gamma^* \, \gamma^*$ cross section.

\begin{figure}[htb]
\centerline{ \DESepsf(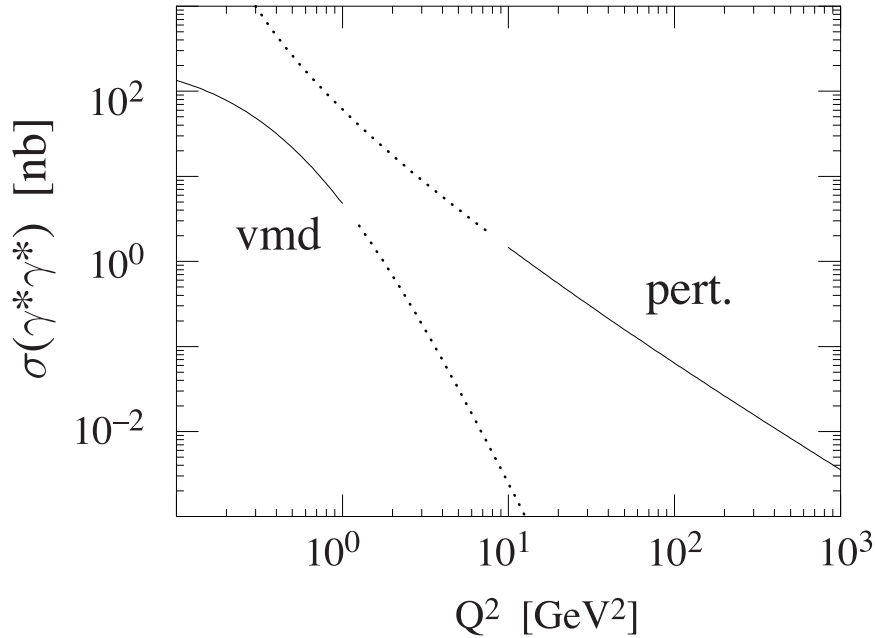 width 13 cm) }
\caption{ $Q^2$-behavior of the vector meson dominance and
perturbative cross sections in lowest order, with
$Q_A^2 = Q_B^2=Q^2$. }
\label{loglog}
\end{figure}

In Fig.~\ref{loglog} we show a log-log plot of the curves
corresponding to the soft and perturbative formulas for the
$Q^2$-behavior of the cross section~\cite{earlier}. For the former,
we use Eqs.~(\ref{ppgap})-(\ref{vmd}), and for the latter we take the
Born approximation to Eq.(\ref{energy}). In this plot we are
interested in emphasizing the dependence of the cross section on
$Q^2$ at fixed $s / Q^2$. For this reason we limit ourselves to the
lowest order formulas, and do not include the higher order summation
of the $\ln(s / Q^2)$ terms that give an enhancement at large
energies. It is understood that such high-energy corrections affect
both the soft and perturbative curves, giving rise to
``soft-pomeron'' and ``hard-pomeron'' effects in the two cases.

The region of intermediate values of $Q^2$ in Fig.~\ref{loglog} ($Q$
of the order of $1 \ {\mbox {GeV}}$) is where the transition from the
soft-scattering regime to the hard-scattering regime is expected to
occur. The mechanism through which this happens is not theoretically
under control at present, and one may consider trying to estimate the
cross section in this region by interpolating between the two curves.

In the next section we will study the prospects for investigating
high energy $ \gamma^* \, \gamma^*$ scattering at
$e^{+} \, e^{-}$ colliders, and we will discuss which regions in
$Q^2$ in Fig.~\ref{loglog} can likely be accessed experimentally at
LEP200 and a future $e^+ e^-$ collider.

\section{Numerical results for the electron-positron cross section}
\label{sec:numerical}

The cross section for high energy virtual photon scattering can be
measured in $e^{+}\, e^{-}$ collisions in which the outgoing leptons
are tagged. The cross section for the electron-positron scattering
process is obtained by folding the $ \gamma^{*} \, \gamma^{*} $ cross
section with the flux of photons from each lepton. Consider the
four-fold differential $e^{+}\, e^{-}$ cross section averaged over
the angle between the lepton scattering planes, Eq.~(\ref{epaave}).
To get an estimate of the rates available to study BFKL effects in
virtual photon scattering at $e^{+}\, e^{-}$ colliders of the present
and next generation, we integrate this cross section over a region
${\cal R}$ determined by cuts that we discuss below:
\begin{equation}
\label{intrate}
\sigma
= \int_{{\cal R}} \, {{d \, x_A} } \,
{{d \, x_B} } \,
{{d \, Q^2_A} \over {Q^2_A}} \,
{{d \, Q^2_B} \over {Q^2_B}} \,
{ { Q^2_A \, Q^2_B \, d \, \sigma^{(e^+ e^-)} } \over
{ d x_A \,
d x_B \,
d Q^2_A \,
d Q^2_B }}
\hspace*{0.4 cm} .
\end{equation}

We choose

i) $Q_A > Q_{\rm min}$, $Q_B > Q_{\rm min}$, where $Q_{\rm min}$ is
a few GeV, in order that the coupling $\alpha_s$ be small, and that
the process be dominated by the perturbative contribution;

ii) $x_A \, x_B \, s_{e e} > {\kappa} \, Q_A \,Q_B$, in order that
the high energy approximation be valid. We discuss the parameter
${\kappa}$ below.

Note that, with these criteria, the photon virtualities $Q^2$ lie in a
range $Q_{\rm min}^2 < Q^2 \ll  s$ in which the equivalent photon
approximation (Eq.~(\ref{epaave})) is expected to work fairly well.
On one hand, kinematical corrections of order $Q^2 / s$ are suppressed
in this range. On the other hand, contributions of order
$m_e^2 / Q^2$ to the $ e \to e \, \gamma$ splitting process can be
neglected.

To choose a value for the parameter ${\kappa}$, we compare the gluon
exchange contribution with contributions that are suppressed by a
power of $s$ \cite{talkfh,earlier}. We consider first the two gluon
exchange graph, for which $\sigma \sim s^0$ for large $s$. Taking the
case $Q_A = Q_B \equiv Q$, we have from Fig.~\ref{s0}
\begin{equation}
\label{sgcomp}
\sigma^{(0)} (s, Q^2) \,
\approx \; {{0.9 \times 16 \,\; \alpha^2 \, \alpha_s^2 \, 
\left( \sum_q
e^2_q \right)^2} \over {Q^2}} \,
\left( 1 + {\cal O} ( Q^2 / s ) \right)
\hspace*{0.8 cm}.
\end{equation}
Next, we consider the leading order (electromagnetic) contribution to
$\gamma^* \, \gamma^* \to q \, {\bar q}$, occurring via quark
exchange, for which $\sigma^{(q)} \sim 1 / s$. In Appendix
\ref{app:quark} we report results for this subprocess. For $Q_A
= Q_B = Q$ and large enough $s$, the corresponding cross section is
well approximated by the formula:
\begin{equation}
\label{sqcomp}
{ \sigma}^{(q)} (s , Q^2) \, \approx \; {{8 \, \pi \, \alpha^2
\, \sum_q e_q^4
\,} \over
{ s }}
\, \left( \ln (s / Q^2) - 1 \right)\,
\left( 1 + {\cal O} ( Q^2 / s ) \right)
\hspace*{0.8 cm} .
\end{equation}
Demanding that the gluon exchange graph give a larger contribution
than the quark exchange graph leads to the requirement 
\begin{equation}
\label{kappalim}
{s \over Q^2} \,
{ 1 \over
 {\left( \ln (s / Q^2) - 1 \right) }} \, \gtrsim \,
\, {1 \over \alpha_s^2} \; \, {{8 \, \pi
\, \sum_q e_q^4
\,} \over {0.9 \times 16 \, \left( \sum_q
e^2_q \right)^2}} \,
\approx \, {1 \over \alpha_s^2}
\hspace*{0.8 cm} .
\end{equation}
For typical perturbative values, $\alpha_s \approx 0.2$, we get $s /
Q^2 \gtrsim 10^2$. We will therefore use $10^2$ as a standard value
for ${\kappa}$.

We note that for $s / Q^2 \gtrsim 10^2$, one is surely entitled to
drop terms in the gluon exchange graphs that are suppressed by powers
of $Q^2/s$, as we have done.

\begin{figure}[htb]
\centerline{ \DESepsf(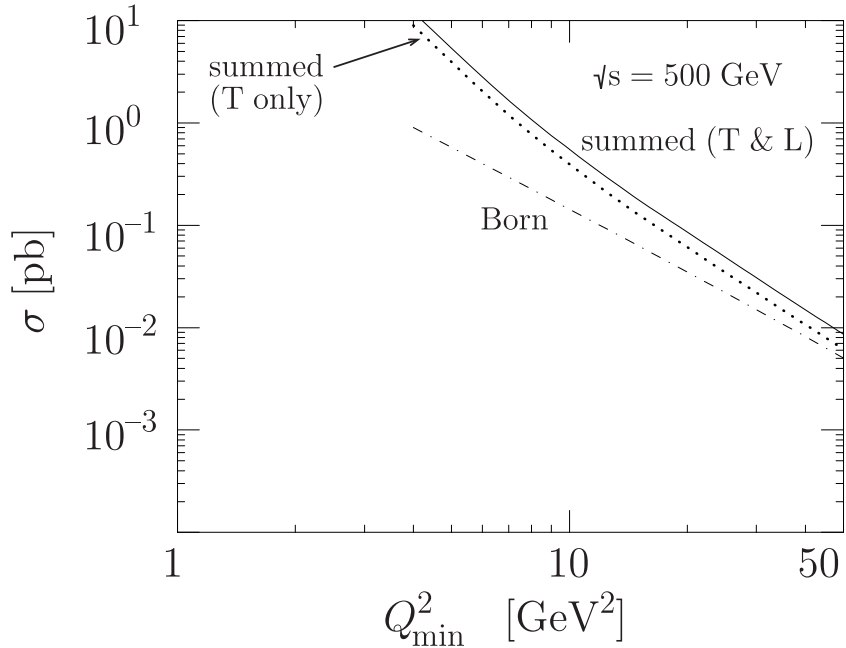 width 13 cm) }
\caption{ The $Q_{\rm min}^2$ dependence of the integrated rate
$\sigma$, Eq.~(\protect\ref{intrate}), for $\protect\sqrt s = 500\
{\rm GeV}$. We take $\kappa = 10^2$. We set the scales $\mu^2$ and
$Q^2$ according to the prescriptions given in
Eqs.~(\protect\ref{cmu}) and (\protect\ref{cqu}). The solid curve
represents the full leading log summation, while the dot-dashed curve
shows the Born result. The dotted curve shows the contribution to the
fully summed result coming from transversely polarized photons. }
\label{f500}
\end{figure}

\begin{figure}[htb]
\centerline{ \DESepsf(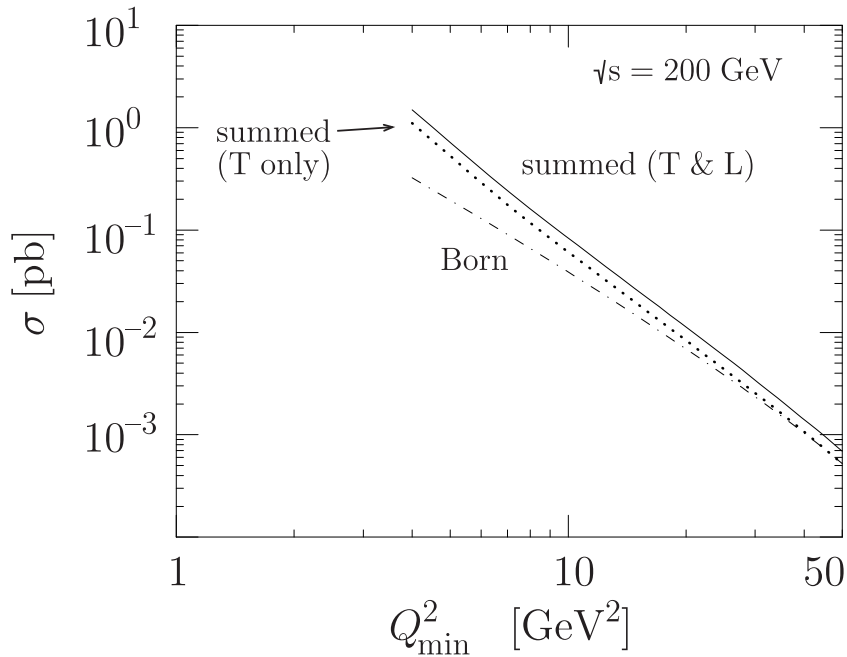 width 13 cm) }
\caption{ Same as in Fig.~\protect\ref{f500} for $\protect\sqrt s =
200\ {\rm GeV}$.}
\label{f200}
\end{figure}

We thus compute the integrated rate $\sigma$ in Eq.~(\ref{intrate})
using the results given in Sec.~\ref{sec:summation} for the
photon-photon cross section, and setting the scales in the running
coupling and in the high energy logarithms according to the
prescriptions discussed in Sec.~\ref{sec:scales}. The dependence
of $\sigma$ on the lower bound $Q_{\rm min}^2$ on the photon
virtualities is shown by the ``summed'' curve in Fig.~\ref{f500} for
the energy of a future $e^+ e^-$ collider.  Fig.~\ref{f200}
shows the cross section for the LEP collider at CERN operating at
$\sqrt s = 200\ {\rm GeV}$. 

The dashed and solid lines in Figs.~\ref{f500} and \ref{f200}
correspond to the result of using, respectively, the Born and the
summed expressions for the photon-photon cross section. At the values
of $\sqrt{s} $ considered in the figures, summation effects enhance
the rates significantly in the range of $Q_{\rm min}$ of a few
${\mbox {GeV}}$. As $Q_{\rm min}$ increases, lowest order
perturbation theory gets closer and closer to the fully summed
prediction, as a result of both $\alpha_s$ becoming small and the
phase space closing up for the high energy logarithms.

In Figs.~\ref{f500} and \ref{f200}, we also plot separately the
contribution to the cross section from purely transverse photons,
that is, the contribution from the term in $\sigma^{(TT)}$ in
Eq.~(\ref{epaave}). We see that this contribution accounts for about
three quarters of the full cross section.

For values of the cuts $Q_{\rm min} = 2 \ {\mbox {GeV}}$, ${\kappa} =
10^2$, we find
\begin{equation}
\label{vallep}
\sigma
\simeq 1.5 \, {\mbox {pb}}
\hspace*{2 cm}
( \sqrt{s} = 200\ {\mbox {GeV}} )
\end{equation}
at LEP200 energies, and
\begin{equation}
\label{valnlc}
\sigma
\simeq 12 \, {\mbox {pb}}
\hspace*{2 cm}
( \sqrt{s} = 500\ {\mbox {GeV}} )
\end{equation}
at the energy of a future collider. These cross sections would give
rise to about $750$ events at LEP200 for a value of the luminosity $L
= 500 \, {\mbox {pb}}^{-1}$, and about $6 \times 10^5$ events at
$\sqrt s = 500\ {\rm GeV}$ for $L = 50 \, {\mbox {fb}}^{-1}$.

The choice of the cuts that can realistically be implemented is
affected by experimental constraints. In particular, the lowest
photon virtualities that can be reached are limited by the angular
acceptance of the detector, according to the relation $Q^2 \approx
(1-x) E^2 \, \theta^2$, where $E$ is the beam energy, $\theta$ the
angle of the tagged lepton, and $x$ is the momentum fraction of the
emitted photon. In this situation, the value
$Q_{\rm min} = 2\ {\mbox {GeV}}$, for which the rates
(\ref{vallep}), (\ref{valnlc}) are given, implies detecting leptons
scattered through angles down to about $20 \ {\mbox {mrad}}$ at
LEP200, which is close to the range of the current luminosity monitors
at the LEP experiments~\cite{lep200angle}.  For a future 500 GeV
collider, $Q_{\rm min} = 2\  {\rm GeV}$ corresponds to a minimum
angle of about $8\ {\mbox {mrad}}$. It appears that working down to
such an angle will be difficult but not impossible~\cite{nlcrep}. If
instead we take $Q_{\rm min} = 6\  {\rm GeV}$, the minimum angle is
$24\ {\mbox {mrad}}$. Then the cross section is about $2 \times
10^{-2}\ {\rm pb}$, corresponding to about $10^3$ events.

\begin{figure}[htb]
\centerline{ \DESepsf(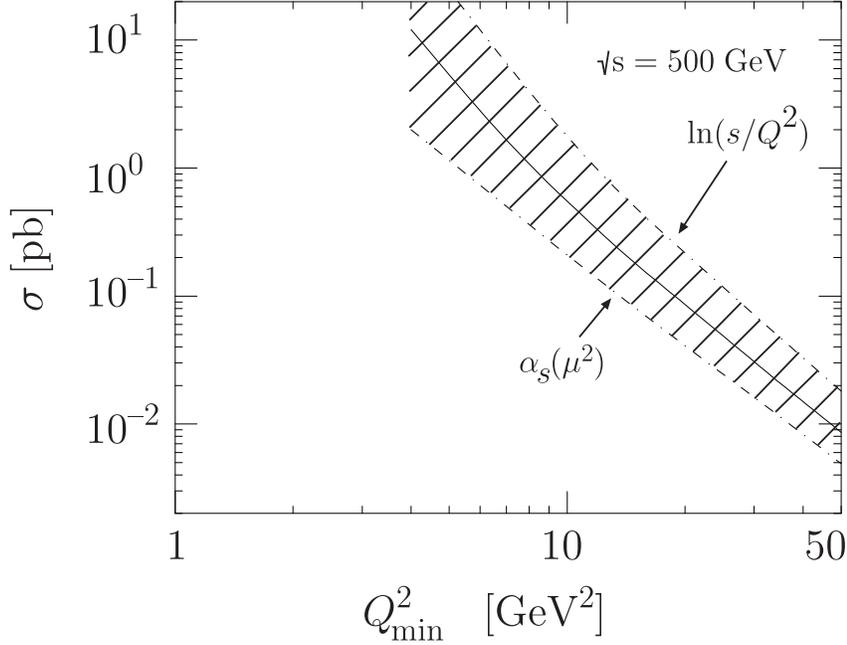 width 13 cm) }
\caption{Estimate of the uncertainty on the leading logarithmic
result for the rate $\sigma$. The solid curve is the summed
result shown in Fig.~\protect\ref{f500}. The dot-dashed curves
summarize the variation of this prediction as a result of
varying the scales in the strong coupling ($\mu^2 \to 4 \mu^2$) and
the high energy logarithms ($Q^2 \to Q^2/4$).
}
\label{band}
\end{figure}

As stated earlier, the numerical results given above depend on
the choice of the scales in $\alpha_s$ and in the high energy
logarithms that enter the photon-photon cross section. For the
calculations described above, we have used the prescriptions given in
Sec.~\ref{sec:scales}. Different scale choices are possible, and
they would affect the predictions at the next-to-leading
logarithmic order, which is beyond the present theoretical accuracy.
We can use the variation of the results with the scale choices to get
an estimate of the uncertainties associated with unknown sub-leading
corrections. We can vary the two scales $\mu^2$ and $Q^2$ (see
Eqs.~(\ref{cmu}), (\ref{cqu})) independently. An illustration of this
is reported in Fig.~\ref{band}. Here we compare the result of
Fig.~\ref{f500} with the curve obtained by multiplying the scale in
$\alpha_s$ by a factor of $4$, $ \mu^2 \to 4 \times \mu^2$, and the
curve obtained by reducing the scale in the high energy logarithms by
a factor of $4$,
$ Q^2 \to Q^2 / 4$. The band between these two curves indicates that
the uncertainty on the leading logarithmic result is fairly large,
and emphasizes the need for improving the accuracy of the
calculations at high energy.

The plots of the cross section versus $Q_{\rm min}^2$ shown in
Figs.~\ref{f500} and \ref{f200} illustrate the expected dependence of 
the photon-photon cross section on the photon virtualities. If we fix
$Q_{\rm min}$ we can look at the dependence on the photon-photon
c.m.\ energy $\sqrt {\hat s}$. It is useful to use
\begin{equation}
\label{hats}
\hat s = x_A\,x_B s
\end{equation}
and the photon-photon rapidity
\begin{equation}
\label{y}
y = {1 \over 2}\,\ln\left( x_A \over x_B\right)
\end{equation}
as variables instead of $x_A$ and $x_B$. Then we define
\begin{equation}
\label{sighats}
{d\sigma \over d \ln\hat s\, dy} \equiv
\int_{Q_{\rm min}}^{\hat s/(\kappa Q_{\rm min})} dQ_A
\int_{Q_{\rm min}}^{\hat s/(\kappa Q_{A})} dQ_B
{d\sigma \over d \ln\hat s\, dy\, dQ_A \, dQ_B}
\end{equation}
Using Eq.~(\ref{epaave}), we can write $d\sigma /( d \ln\hat s\, dy)$
as
\begin{eqnarray}
{d\sigma \over d \ln\hat s\, dy}&=&
\left({ \alpha \over \pi}\right)^{\!2} 
x_A P^{(T)}_{\gamma/e^+}(x_A)\
x_B P^{(T)}_{\gamma/e^-}(x_B)
\nonumber\\
&&\times
\int_{Q_{\rm min}}^{\hat s/(\kappa Q_{\rm min})} {dQ_A \over Q_A}
\int_{Q_{\rm min}}^{\hat s/(\kappa Q_{A})} {dQ_B \over Q_B}\
\sigma_{\gamma^*\gamma^*}^{TT}(\hat s, Q_A^2, Q_B^2)
\nonumber\\
&& + \cdots \,,
\end{eqnarray}
where we omit three similar terms. We see that $d\sigma /( d \ln\hat
s\, dy)$ is very directly related to the $\gamma\,\gamma$ cross
section.

\begin{figure}[htb]
\centerline{\DESepsf(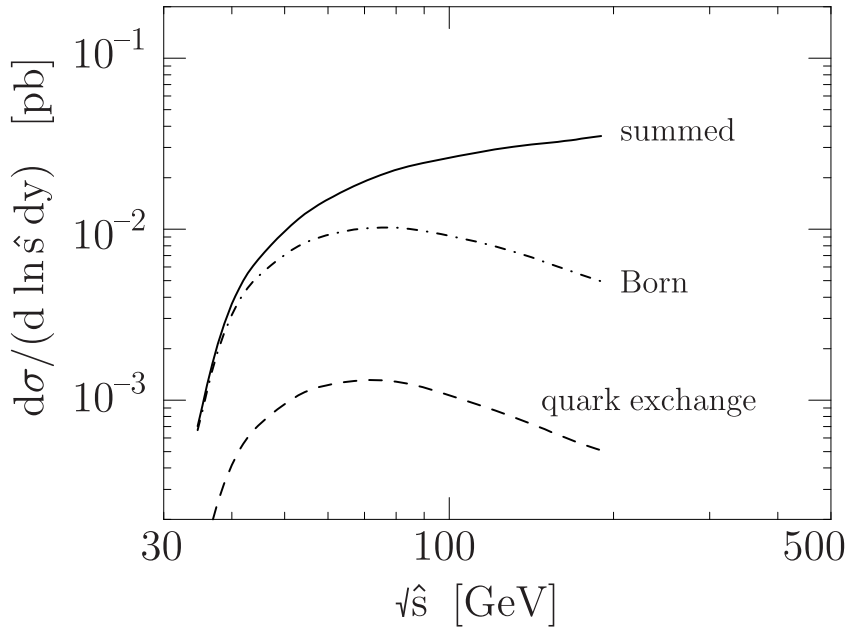 width 13 cm)}
\caption{The cross section $d\sigma / (d \ln\hat s\, dy)$,
Eq.~(\protect\ref{sighats}), at $y = 0$ for $\protect\sqrt s = 200\
{\rm GeV}$. The solid curve is the summed BFKL result. The dot-dashed
curve is the Born result. The dashed curve shows the cross section
arising from the scattering of (transversely polarized) photons via
quark exchange. The cuts are $Q_{\rm min}^2 = 10\ {\rm GeV}^2$ and
$\kappa = 10^2$.}
\label{hats200}
\end{figure}

\begin{figure}[htb]
\centerline{\DESepsf(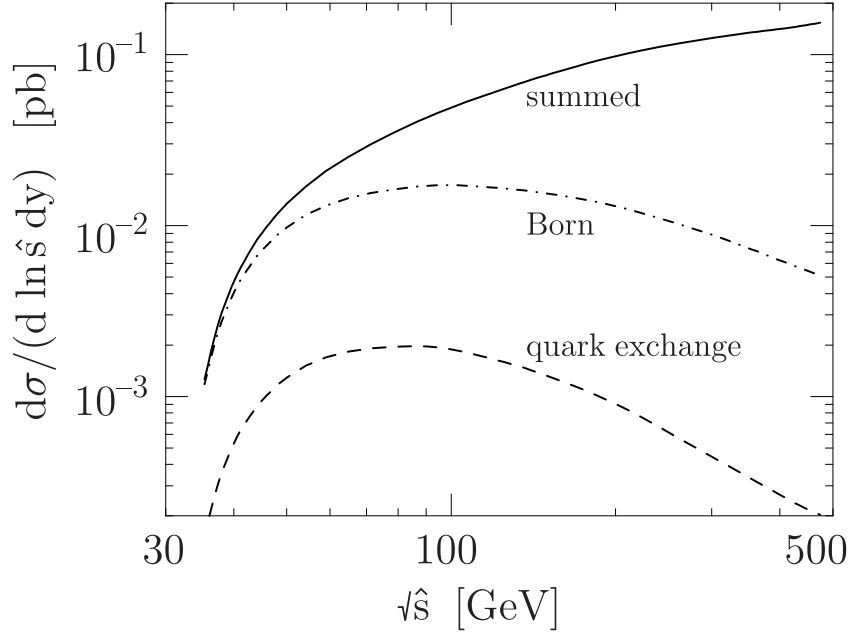 width 13 cm)}
\caption{Same as in Fig.~\protect\ref{hats200} but for
$\protect\sqrt s = 500\ {\rm GeV}$. }
\label{hats500}
\end{figure}

We plot ${d\sigma / (d \ln\hat s\, dy)}$ at $y = 0$ in
Fig.~\ref{hats200} for $\sqrt{s} = 200\ {\rm GeV}$ and in
Fig.~\ref{hats500} for $\sqrt{s} = 500\ {\rm GeV}$. Here we choose
$Q_{\rm min}^2 = 10\ {\rm GeV}^2$ and $\kappa = 10^2$. In each case,
we show a curve for the Born level cross section and another for the
full BFKL cross section. We also show the cross section arising from
the scattering of (transversely polarized) photons via quark exchange
instead of gluon exchange. We see that, with our choice of cuts,
quark exchange scattering is suppressed.

For $\sqrt{\hat s} \lesssim 50\ {\rm GeV}$ the cross section shows a
strong dependence on the cut $\kappa\, Q_A Q_B < \hat s$. With our
choice of $\kappa = 100$ and with $Q_A, Q_B > Q_{\rm min} =
\sqrt {10} \ {\rm GeV}$, the cross section is forced to vanish for
$\sqrt{\hat s} < 31.6\ {\rm GeV}$. As $\sqrt{\hat s}$ increases, the
effect of this cut on the $Q_A$ and $Q_B$ integrations becomes less
and less important. Since $\sigma_{\gamma^*\gamma^*}(\hat s, Q_A^2,
Q_B^2)$ is independent of $\hat s$ at Born level, the cross section
begins to flatten out as $\sqrt{\hat s}$ increases to about $100 \
{\rm GeV}$. For larger values of $\sqrt{\hat s}$, the Born cross
section decreases because of the influence of the photon flux factor
$x_A P(x_A)\ x_B P(x_B) = (\hat s /s) P([{\hat s /s}]^{1/2})^2$. For
the summed BFKL curve, the growth of $\sigma_{\gamma^*\gamma^*}(\hat
s, Q_A^2, Q_B^2)$ overcomes the effect of the photon flux factor, so
that the cross section rises with $\sqrt{\hat s}$.

The curves for $\sqrt{s} = 200\ {\rm GeV}$ and  $\sqrt{s} = 500\ {\rm
GeV}$ are similar. The main difference is that at $\sqrt{s} = 500\
{\rm GeV}$ there is more available range for $\sqrt{\hat s}$.

Our plots are for $y = 0$. The available range of $y$ is $|y| <
{\scriptstyle{1\over 2}}\,\ln(s/\hat s)$. Thus the cross section
integrated over $y$ goes to zero as $\sqrt{\hat s} \to \sqrt s$.

We see from the results presented above that at a future $e^+ e^-$
collider it should be possible to probe the effects of pomeron
exchange in a range of $Q^2$ where summed perturbation theory
applies. One should be able to investigate this region in detail by
varying $Q_A$, $Q_B$ and ${\hat s} = x_A \, x_B \, s_{ e e }$
independently. At LEP200 such studies appear to be more problematic
mainly because of limitations in luminosity. Even with a modest
luminosity, however, one can access the region of relatively low
$Q^2$ in the graph of Fig.~\ref{loglog} if one can get down to small
enough angles. This would allow one to examine experimentally the
transition between soft and hard scattering.

We now move on to the angular distribution for the $e^+ \, e^-$
scattering cross section, and consider the asymmetries $A_1$, $A_2$
introduced in Eq.~(\ref{dsdphi}). As pointed out in
Sec.~\ref{sec:polarization}, $A_1$ is zero at leading order. On the
other hand, $A_2$ is given by an equivalent-photon formula in terms of
the asymmetry ${\cal A}$ for the $\gamma^* \, \gamma^*$ scattering
process discussed in Sec.~\ref{subsec:asym}. This reads
\begin{eqnarray}
\label{epaasy}
&& A_2
\, { { Q^2_A \, Q^2_B \, d \, \sigma^{(e^+ e^-)} } \over
{ d x_A \,
d x_B \,
d Q^2_A \,
d Q^2_B }}
\nonumber \\
&& =
\left( {\alpha \over {2 \, \pi}} \right)^2 \,
{{1 - x_A } \over x_A} \,
{{1 - x_B } \over x_B} \,
{\cal A}_{\gamma^* \, \gamma^*} (x_A \, x_B \, s, Q^2_A , Q^2_B )
{\overline {\sigma}}_{\gamma^* \, \gamma^* }
(x_A \, x_B \, s, Q^2_A , Q^2_B )
\,\;\;\;,
\end{eqnarray}
with ${\cal A}$ and ${\overline {\sigma}}$ being given in
Eqs.~(\ref{energyasy}), (\ref{energy}).

We can use the same cuts discussed earlier in this section to
integrate Eq.~(\ref{epaasy}), and thus define
\begin{equation}
\label{intrateasy}
{\tilde A}_2
= {1 \over \sigma} \, \int_{{\cal R}} \, {{d \, x_A} } \,
{{d \, x_B} } \,
{{d \, Q^2_A} \over {Q^2_A}} \,
{{d \, Q^2_B} \over {Q^2_B}} \, A_2 \,
{ { Q^2_A \, Q^2_B \, d \, \sigma^{(e^+ e^-)} } \over
{ d x_A \,
d x_B \,
d Q^2_A \,
d Q^2_B }}
\hspace*{0.4 cm} ,
\end{equation}
where $\sigma$ is the integrated rate in Eq.~(\ref{intrate}).
Performing the integral numerically, we find that the asymmetry
$ {\tilde A}_2$ is very small. As noted in Sec.~\ref{sec:summation},
the role of the summed BFKL terms is that of reducing the magnitude
of the asymmetry with respect to the Born order result. At a
500 GeV collider, in the range of the angular and energy cuts
previously described, we find $ {\tilde A}_2 \simeq 
10^{-3}$. We observe that spin effects in photon-photon scattering at
high energies are interesting, but the predicted asymmetries are
either zero or small.

\section{Conclusions}
\label{sec:conclusions}

Understanding the behavior of high energy hadron reactions from a
fundamental perspective within QCD is an important goal of particle
physics. As we have shown in this paper, virtual photon scattering
$\gamma^*(Q_A^2) + \gamma^*(Q_B^2) \to hadrons $ at high energies, $s
>> Q_A^2, Q_B^2$, provides a remarkable window into pomeron physics.
The total cross section can be studied as a function of the
space-like mass of each incident projectile. Most importantly, the
process can be investigated in the regime where the photons both have
large virtuality, so that one can use the framework of perturbative
QCD.

Compared to tests of the QCD pomeron behavior based on deeply
inelastic structure functions, the measurement of the total cross
section for sufficiently off-shell photons is free from the
long-distance ambiguities related to the structure of the hadronic
target. On the other hand, unlike tests based on associated jet
production in lepton-hadron or hadron-hadron collisions, the
$\gamma^* \gamma^*$ measurement is fully inclusive and therefore it
does not depend on specifying the details of the final state.

The scattering of highly virtual photons can be described as the
interaction of two incident color singlet $q \bar q$ pairs of small
transverse size interacting through multiple gluon exchange. We have
studied this reaction both in the Born approximation (corresponding
to two-gluon exchange) and also with the inclusion of the
higher-order summation encompassed by the BFKL equation. The cross
section at high energies and large virtuality takes a factorized form
in transverse coordinates. However, it does not factorize simply into
separate functions of $Q_A^2$ and $Q_B^2$, which reflects the
cut structure of the BFKL pomeron in the complex angular momentum
plane. We have also examined the background contribution from quark
exchange, a process which is power-law suppressed at high energy.

According to this analysis, the $\gamma^* \gamma^* $ cross section
falls off at high virtuality only as $1/Q^2$, where $Q^2 \sim {\mbox
{max}} \{ Q_A^2 , Q_B^2\} $. The rate for sensitive tagged-lepton
experiments at high energy $e^{\pm}e^-$ or $\mu^{\pm}\mu^-$ colliders
is thus not negligible. In particular it appears that the main
features of the perturbative QCD predictions, such as the energy
dependence, the factorization properties of the cross section, the
scaling laws in $Q_A^2$, $Q_B^2$, as well as the polarization and
azimuthal correlations can be tested in detail at a high-energy and
high-luminosity next linear collider. We have also found that an
interesting first look at virtual photon scattering can be obtained
from the tagged lepton events measured in the luminosity monitors of
present experiments at LEP200.

More precisely, we estimate that, in the region of photon
virtualities where summed perturbation theory is expected to apply,
there should be several hundred events at LEP200, and about $10^5$
events at a future 500 GeV collider with an integrated luminosity of
$50\ {\rm fb}^{-1}$. We also find that the enhancement due to BFKL
pomeron terms over the Born cross section is sizable, and should be
visible particularly in the ${\hat s}$-distribution of the cross
section, with ${\hat s} = x_A \, x_B \, s_{ e e }$.

The dependence of the cross section on the photon virtualities
$Q_A$ and $Q_B$ is perturbative, and can be predicted in the
framework of the BFKL equation. These predictions can be tested by
measuring the angles of the recoil leptons. Both the case in which
the two photon virtualities are varied together ($Q_A \sim Q_B$) and
the case in which they are kept far apart ($Q_A \gg Q_B$) are of
interest. In the second case one gets to observe the structure
function of a virtual photon at small Bjorken-$x$.

The spin structure is rich, but hard to observe. Most of the
observable cross section comes from the scattering of two
transversely polarized photons. For this part of the cross section,
there is an asymmetry in the angular distribution of  the outgoing
leptons, but this asymmetry is  less than $1 \%$.

In the region of low photon virtualities ($Q_A , \, Q_B $ smaller
than a few ${\mbox {GeV}}$), the photon-photon cross section becomes
dominated by soft interactions. Here one cannot use a perturbative
analysis. On the other hand, one may explore experimentally at what
scales the breakdown of the perturbative result occurs, and how this
is connected to the onset of the phenomenological ``soft-pomeron''
behavior.

The theory that is available at present is leading logarithmic and
therefore is affected by rather large uncertainties. These
uncertainties can be parametrized in terms of two mass scales, the
transverse scale that controls the running coupling and the
longitudinal scale associated with the high energy logarithms. A
next-to-leading logarithmic calculation would help determine
these scales. Such a calculation could make the theoretical
predictions much more precise. At the largest values of $\hat s$, new
effects related to unitarity and diffusion may become important. If
so, an improved theory that deals with these effects would be testable
at a future $e^+ e^-$ collider.

\acknowledgments

We are grateful to J.\ Bjorken and A.\ Mueller for discussions and for
their interest in this work. We thank D.\ Strom for useful advice.
This work was supported in part by the United States Department of
Energy grants DE-AC03-76SF00515 and DE-FG03-96ER40969.

\appendix

\section{The Born order calculation}
\label{app:born}

We start with the expression (\ref{m1}) for the amplitude
corresponding to the graph in Fig.~\ref{ftwoglu}. The overall charge
factor in Eq.~(\ref{m1}) is
\begin{equation}
\label{charge}
\sum_{ a , b} \, g_s^4 \, e^2_a \, e^2_b \, e^4 \,
{\mbox {\rm Tr}} \left( t^r t^s \right) \,
{\mbox {\rm Tr}} \left( t^r t^s \right) =
32 \, \alpha^2 \, \alpha_s^2 \, \left( \sum_q e^2_q \right)^2 \,
(2 \, \pi)^4 \;\;\;,
\end{equation}
where we have used the color trace ${\mbox {\rm Tr}} \left( t^r t^s
\right) = (1 / 2) \delta^{ r s} $.

We use the mass shell constraints on the final quarks, $p_A$
and $p_B$, to eliminate the integrals over their ``$-$'' and ``$+$''
components, respectively, thus obtaining
\begin{equation}
\label{xaprime}
\int {{d^4 p_A} \over {(2 \, \pi)^4}}
\; 2 \, \pi \, \delta_{+}\!\left( p_A^2 \right)
\to \int {{d \, z_A} \over {2 \, z_A}} \, d^2 \, {{\bf p}}_A \,
(2 \, \pi)^{-3} \;\;, \;\;\;
z_A^{\prime} = {{{\bf p}}_A^2 \over {2 \; z_A \; q^{+}_A q^{-}_B}}
\;\;,
\end{equation}
\begin{equation}
\label{xbprime}
\int {{d^4 p_B} \over {(2 \, \pi)^4}}
\; 2 \, \pi \, \delta_{+}\!\left( p_B^2 \right)
\to \int {{d \, z_B} \over {2 \, z_B}} \, d^2 \, {{\bf p}}_B \,
(2 \, \pi)^{-3} \;\;, \;\;\;
z_B^{\prime} 
= {{{\bf p}}_B^2 \over {2 \; z_B \; q^{+}_A q^{-}_B}} \;\;.
\end{equation}
We use the mass shell constraints on the antiquarks to eliminate the
integrals over the ``$-$'' and ``$+$'' components of the exchanged
momentum $k$, as follows:
\begin{eqnarray}
\label{kminus}
&&\int {{d \, k^{-}} \over {2 \, \pi}}
\; 2 \, \pi \, \delta_{+}\!\left( (q_A-p_A-k)^2 \right) \to
{1 \over {2 \; (1 - z_A) \; q^{+}_A}} \;\;, \,
\nonumber\\&&
k^{-} \simeq - {1 \over {2 \, q_A^{+}}} \,
\left( Q_A^2 + { { {{\bf p}}_A^2 } \over z_A} +
{ { ({{\bf p}}_A + {{\bf k}})^2 } \over {1 - z_A}} \right) \;\;\;,
\end{eqnarray}
\begin{eqnarray}
\label{kplus}
&&\int {{d \, k^{+}} \over {2 \, \pi}}
\; 2 \, \pi \, \delta_{+}\!\left( (q_B-p_B+k)^2 \right) \to
{1 \over {2 \; (1 - z_B) \; q^{-}_B}} \;\;,
\nonumber\\&&
k^{+} \simeq {1 \over {2 \, q_B^{-}}} \,
\left( Q_B^2 + { { {{\bf p}}_B^2 } \over z_B} +
{ { ({{\bf p}}_B - {{\bf k}})^2 } \over {1 - z_B}} \right) \;\;\;.
\end{eqnarray}
Note that in Eqs.~(\ref{kminus}), (\ref{kplus}) we have neglected
terms of order $k^{+} / \sqrt{s}$ and $k^{-} / \sqrt{s}$ with respect
to unity, consistently with the high energy approximation.

Eq.~(\ref{m1}) can then be rewritten as
\begin{eqnarray}
\label{m2}
&~& {|{\cal M} |}^2 =
32 \, \alpha^2 \, \alpha_s^2 \, \left( \sum_q e^2_q \right)^2 \,
(2 \, \pi)^4 \,
\int \; {{d^2 \, {\bf k}} \over
{(2 \, \pi)^2}} \;
{{d^2 \, {{\bf p}}_A} \over {(2 \, \pi)^2}} \;
{{d^2 \, {{\bf p}}_B} \over {(2 \, \pi)^2}}
\nonumber
\\
&~& \times \,
\int_0^1 {1 \over {2 \, \pi}} \;
{{d \, z_A} \over {2 \, z_A \, ( 1 - z_A)}}
\int_0^1 {1 \over {2 \, \pi}} \;
{{d \, z_B} \over {2 \, z_B \, ( 1 - z_B)}}
\; {1 \over {2 \, s}} \,
{ 1 \over {(k^2)^2} } \,
\nonumber \\
&~& \times \, { {
{\mbox {\rm Tr}} \left[ \,
{\rlap/p}_A \, \gamma_\alpha \,
( {\rlap/p}_A + \rlap/k ) \, {\rlap/\varepsilon}_A \,
({\rlap/q}_A - {\rlap/p}_A - \rlap/k ) \,
\gamma_\beta \,
( {\rlap/p}_A - {\rlap/q}_A ) \, {\rlap/\varepsilon}_A
\right]
} \over {
\left( (p_A + k)^2 + i \, \varepsilon \right)
\left( (p_A - q_A)^2 - i \, \varepsilon \right) } } \,
\nonumber \\
&~& \times \, { {
{\mbox {\rm Tr}} \left[ \,
{\rlap/p}_B \, \gamma^\alpha \,
( {\rlap/p}_B - \rlap/k ) \, {\rlap/\varepsilon}_B \,
({\rlap/q}_B - {\rlap/p}_B + \rlap/k ) \,
\gamma^\beta \,
( {\rlap/p}_B - {\rlap/q}_B ) \, {\rlap/\varepsilon}_B
\right]
} \over {
\left( (p_B - k)^2 + i \, \varepsilon \right)
\left( (p_B - q_B)^2 - i \, \varepsilon \right) } } \;\;.
\end{eqnarray}

We now re-express the denominators and numerators of the amplitude
(\ref{m1}) in the high energy limit. With the neglect of terms of
order $k^{+} / \sqrt{s}$ and $k^{-} / \sqrt{s}$, the denominators
take the form
\begin{equation}
\label{denoa1}
(p_A+k)^2 \simeq - {1 \over {1-z_A}} \, \left( z_A \, (1-z_A) \, Q^2_A
+ ({{\bf p}}_A + {{\bf k}})^2 \right) \,\;, \;\;\;
\end{equation}
\begin{equation}
\label{denoa2}
(p_A-q_A)^2 \simeq - {1 \over z_A} \, \left( z_A \, (1-z_A) \, Q^2_A
+ {{\bf p}}_A^2 \right) \;\;\;,
\end{equation}
and analogously
\begin{equation}
\label{denob1}
(p_B-k)^2 \simeq - {1 \over {1-z_B}} \, \left( z_B \, (1-z_B) \, Q^2_B
+ ({{\bf p}}_B - {{\bf k}})^2 \right) \,\;, \;\;\;
\end{equation}
\begin{equation}
\label{denob2}
(p_B-q_B)^2 \simeq - {1 \over z_B} \, \left( z_B \, (1-z_B) \, Q^2_B
+ {{\bf p}}_B^2 \right) \;\;\;.
\end{equation}
In the numerator, light-cone gluon polarizations are dominant at high
energy, and therefore we are led to calculate the product of traces
\begin{eqnarray}
\label{trplusminus}
&~& T_{A}^{+ +} \, T_B^{- -}
\\
&~& =
{\mbox {\rm Tr}} \left[ \,
{\rlap/p}_A \, \gamma^{+} \,
( {\rlap/p}_A + \rlap/k ) \, {\rlap/\varepsilon}_A \,
({\rlap/q}_A - {\rlap/p}_A - \rlap/k ) \,
\gamma^{+} \,
( {\rlap/p}_A - {\rlap/q}_A ) \, {\rlap/\varepsilon}_A
\right] \;
\nonumber \\
&~& \times \, {\mbox {\rm Tr}} \left[ \,
{\rlap/p}_B \, \gamma^{-} \,
( {\rlap/p}_B - \rlap/k ) \, {\rlap/\varepsilon}_B \,
({\rlap/q}_B - {\rlap/p}_B + \rlap/k ) \,
\gamma^{-} \,
( {\rlap/p}_B - {\rlap/q}_B ) \, {\rlap/\varepsilon}_B
\right] \;.
\nonumber
\end{eqnarray}
The result reads
\begin{eqnarray}
\label{trres}
&~& T_{A}^{ + +} \, T_{B}^{ - -}
\\
&=&
8 \, (q_A^{+})^2 \, \left( 4 z_A (1 - z_A)
{\bf{\varepsilon}}_A \cdot {{\bf p}}_A \;
{\bf{\varepsilon}}_A \cdot ({{\bf p}}_A + {\bf k}) -
{{\bf p}}_A \cdot ({{\bf p}}_A + {\bf k}) \right) \,
\nonumber \\
&\times& 8 \, (q_B^{-})^2 \, \left( 4 z_B (1 - z_B)
{\bf{\varepsilon}}_B \cdot {{\bf p}}_B \;
{\bf{\varepsilon}}_B \cdot ({{\bf p}}_B - {\bf k}) -
{{\bf p}}_B \cdot ({{\bf p}}_B - {\bf k}) \right) \;\;\;.
\nonumber
\end{eqnarray}

Substituting Eqs.~(\ref{denoa1})-(\ref{denob2}) and (\ref{trres}) in
Eq.~(\ref{m2}), we obtain the expression (\ref{m3}) for the amplitude
corresponding to the graph in Fig.~\ref{ftwoglu}.

The total cross section is arrived at by adding the contributions
from the other graphs according to the replacements described in the
text below Eq.~(\ref{m3}), and dividing by $ 2 \, s$. The result
reads
\begin{eqnarray}
\label{m4}
&~& \sigma^{(0)}_{\gamma^* \, \gamma^*} =
128 \, \alpha^2 \, \alpha_s^2
\, \left( \sum_q e^2_q \right)^2
\, (2 \, \pi)^2 \,
\int \; {{d^2 \, {\bf k}} \over {(2 \, \pi)^2}} \;
{{d^2 \, {{\bf p}}_A} \over {(2 \, \pi)^2}} \;
{{d^2 \, {{\bf p}}_B} \over {(2 \, \pi)^2}} \;
\int_0^1
{d \, z_A}
\int_0^1
{d \, z_B}
\, {1 \over {({{\bf k}}^2)^2} }
\\
&~& \times \, \left\{ { {
\left[ 4 z_A (1 - z_A)
{\bf{\varepsilon}}_A \cdot {{\bf p}}_A \;
{\bf{\varepsilon}}_A \cdot ({{\bf p}}_A + {\bf k}) -
{{\bf p}}_A \cdot ({{\bf p}}_A + {\bf k}) \right]
} \over {
\left[ z_A \, (1-z_A) \, Q^2_A
+ ({{\bf p}}_A + {{\bf k}})^2 \right] \,
\left[ z_A \, (1-z_A) \, Q^2_A
+ {{\bf p}}_A^2 \right]
} } \right.
\nonumber
\\&&\hspace*{1cm} \left.
-
{ {
\left[ 4 z_A (1 - z_A)
({\bf{\varepsilon}}_A \cdot {{\bf p}}_A )^2\;
- {{\bf p}}_A^2 \right]
} \over {
\left[ z_A \, (1-z_A) \, Q^2_A
+ {{\bf p}}_A^2 \right]^2
} }
\right\}
\nonumber \\
&~& \times \, \left\{ { {
\left[ 4 z_B (1 - z_B)
{\bf{\varepsilon}}_B \cdot {{\bf p}}_B \;
{\bf{\varepsilon}}_B \cdot ({{\bf p}}_B - {\bf k}) -
{{\bf p}}_B \cdot ({{\bf p}}_B - {\bf k}) \right]
} \over {
\left[ z_B \, (1-z_B) \, Q^2_B
+ ({{\bf p}}_B - {{\bf k}})^2 \right] \,
\left[ z_B \, (1-z_B) \, Q^2_B
+ {{\bf p}}_B^2 \right]
} } \right.
\nonumber
\\&&\hspace*{1cm} \left.
-
{ {
\left[ 4 z_B (1 - z_B)
({\bf{\varepsilon}}_B \cdot {{\bf p}}_B )^2\;
- {{\bf p}}_B^2 \right]
} \over {
\left[ z_B \, (1-z_B) \, Q^2_B
+ {{\bf p}}_B^2 \right]^2
} }
\right\}
.
\nonumber
\end{eqnarray}
This coincides with Eq.~(\ref{convsigma}) in the text once the
explicit expression (\ref{capitalg}) for $G$ is used.

Using Eqs.~(\ref{m2})-(\ref{trplusminus}), the tensor $ {\cal G}^{\,
\mu \, \nu \,} ({\bf k}; Q^2)$ introduced in Eq.~(\ref{curlyg}) takes
the form
\begin{eqnarray}
\label{curlygex}
&& {\cal G}^{\, \mu \, \nu \,} ({\bf k}; Q^2) =
{\alpha \, \alpha_s \over 4 (q_A^+)^2}
 \, \left( \sum_q e^2_q \right) \,
\int \; {{d^2 \, {{\bf p}}} \over {\pi}} \;
\int_0^1 \,
{d \, z}
\,
\\
&& \times \, \left\{
{ { {\mbox {\rm Tr}} \left[ \,
{\rlap/p} \, \gamma^{+} \,
( {\rlap/p} + \rlap/k ) \, \gamma^{\mu} \,
({\rlap/q} - {\rlap/p} - \rlap/k ) \,
\gamma^{+} \,
( {\rlap/p} - {\rlap/q} ) \, \gamma^{\nu}
\right] }
\over {
\left[
({{\bf p}} + {{\bf k}})^2 + z \, (1-z) \, Q^2 \right] \,
\left[
{{\bf p}}^2 + z \, (1-z) \, Q^2 \right]
} }
+ \, symm. \, \right\}
\;\;\;,
\nonumber
\end{eqnarray}
where the additive symmetric terms are obtained from the replacements
given in Sec.~\ref{sec:notations}.

\section{Comparison with the cross section from quark exchange }
\label{app:quark}

The gluon exchange diagrams discussed in the text provide the dominant
contribution to the photon-photon cross section in the high energy
limit. They give rise to constant (in Born order) or logarithmic (in
higher orders) terms at large $s$ in the cross section. This appendix
is concerned with quark exchange contributions, which vanish in the
large energy limit. We examine quark exchange in order to estimate
the energy at which gluon exchange becomes dominant.

\begin{figure}[htb]
\centerline{ \DESepsf(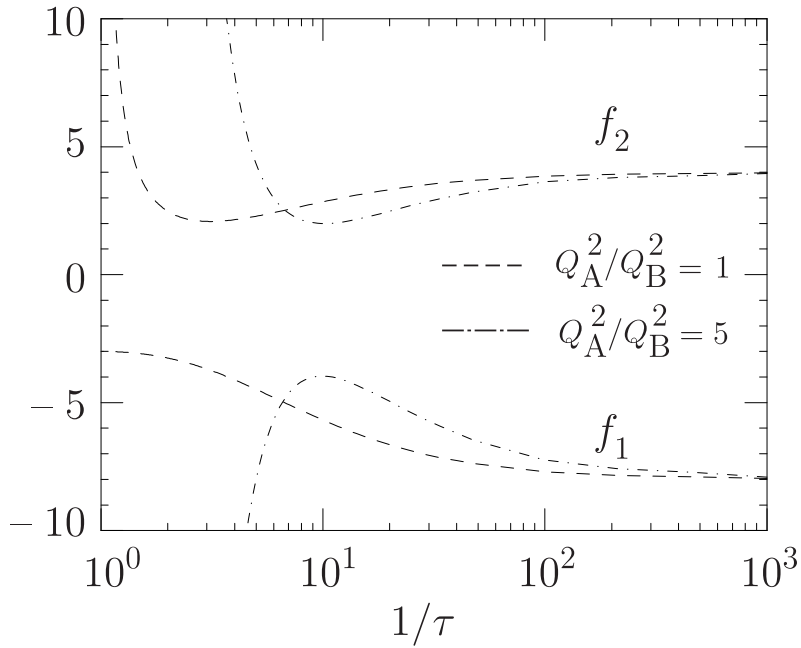 width 13 cm) }
\caption{The $\tau$-dependence of the functions $f_1$ and $f_2$ that
enter the leading order expression for ${ \sigma}^{(q)}$,
Eq.~(\protect\ref{sqexunpol}). We report $f_1$ and $f_2$ for two
different values of $Q_A^2 / Q_B^2$. }
\label{qfun}
\end{figure}

The leading-order term of quark-exchange type comes from the purely
electromagnetic process
\begin{equation}
\label{qexreac}
\gamma^{*} (q_A) + \gamma^{*} (q_B) \to
q (p) + {\bar q} ({\bar p}) \;\;\; .
\end{equation}
This contribution is suppressed by a power of $s$ at high energies,
$ \sigma^{(q)} \sim 1 / s $. To express the cross section for this
process, we parametrize the incoming photon momenta as in
Eq.~(\ref{qaqb}), and introduce the variables
\begin{equation}
\label{xiab}
\xi_A = {{Q_A^2} \over { 2 \, q^{+}_A \, q^{-}_B }} \hspace*{0.7 cm} ,
\hspace*{1 cm}
\xi_B = {{Q_B^2} \over { 2 \, q^{+}_A \, q^{-}_B }} \hspace*{1 cm} ,
\end{equation}
in terms of which the total energy $s$ has the expression
\begin{equation}
\label{sxi}
s = 2 \, q^{+}_A \, q^{-}_B \,\;\;
(1 - \xi_A) \, (1 - \xi_B) \hspace*{1 cm} .
\end{equation}
In the high energy region one has
$
\xi_A , \, \xi_B \ll 1$, and
$s \sim 2 \, q^{+}_A \, q^{-}_B $ as in Eq.~(\ref{qaplusqbminus}). 

\begin{figure}[htb]
\centerline{ \DESepsf(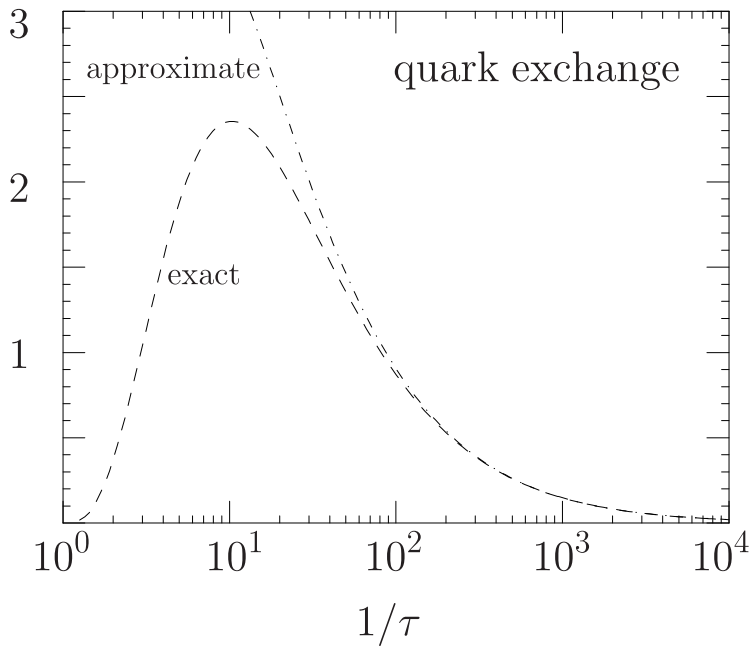 width 13 cm) }
\caption{The $\tau$-dependence of the quark-exchange contribution to
the $\gamma^* \gamma^*$ cross section in leading order. We take
$Q_A = Q_B$, and we plot the rescaled cross section $Q_A \, Q_B \,
\sigma^{(q)} / \left( \alpha^2 \, ( \sum_q e_q^4 ) \right) $. The
dashed curve is the exact expression, Eq.~(\protect\ref{sqexunpol}),
while the dot-dashed curve is the expression approximated for high
energies, Eq.~(\protect\ref{sqasy}). }
\label{qsig}
\end{figure}

The cross section for the process (\ref{qexreac}) to order
$\alpha^2$, averaged over the transverse photon polarizations, has
the form
\begin{equation}
\label{sqexunpol}
{ \sigma}^{(q)} = {{2 \, \pi \, \alpha^2
\, \sum_q e_q^4
\,} \over
{ Q_A \, Q_B }} {{\sqrt{\xi_A \, \xi_B}} \over
{2 \, (1 - \xi_A \, \xi_B)}}
\left[
{ { f_1}
} +
{{ f_2
\, \ln \left( 1/(\xi_A \, \xi_B) \right)}
}
\right]
\hspace*{0.8 cm} ,
\end{equation}
where $f_1$ and $f_2$ are rational functions of $\xi_A , \xi_B$, and
are plotted in Fig.~\ref{qfun} versus the variable $\tau = 
\sqrt{\xi_A\, \xi_B}$ for different values of the ratio $\rho = \xi_A
/ \xi_B$.

The $\tau$-dependence of the cross section (\ref{sqexunpol}) is
reported in Fig.~\ref{qsig} for the case of equal virtualities. The
cross section vanishes at the kinematic threshold $\tau = 1$, it has
a maximum around $\tau \sim 10^{-1}$, then it falls off and vanishes
for $\tau \to 0$ (corresponding to high energy) like $\tau \, \ln
\tau$, according to the asymptotic formula
\begin{equation}
\label{sqasy}
{ \sigma}^{(q)} \simeq {{2 \, \pi \, \alpha^2
\, \sum_q e_q^4
\,} \over
{ Q_A \, Q_B }}
\, 4 \, \tau \, \left( \ln (1 / \tau) - 1 \right)
\hspace*{0.5 cm} , \hspace*{0.8 cm} \tau \ll 1
\hspace*{0.8 cm} .
\end{equation}
The power suppression with $\tau$ at small $\tau$ is the one expected
from the exchange of a spin-$1 / 2$ line in the $t$-channel. The
logarithmic enhancement is associated with the integration over the
region of small angles at the splitting vertex $\gamma^* \to q \,
{\bar q}$ in the limit of small photon virtuality. The behavior of the
cross section is qualitatively the same in the case of unequal
virtualities.


\begin{references}   
\bibitem{BFKL}
      L.N.\ Lipatov, 
          Sov.\ J.\ Nucl.\ Phys.\ {\bf 23}, 338 (1976); 
      E.A.\ Kuraev, L.N.\ Lipatov and V.S.\ Fadin, 
          Sov.\ Phys.\ JETP  {\bf 45}, 199 (1977) ;
      I.\ Balitskii and L.N.\ Lipatov,
          Sov.\ J.\ Nucl.\ Phys.\ {\bf 28}, 822 (1978).
\bibitem{abra}
      H.\ Abramowicz, plenary talk at ICHEP96 (Warsaw, July 1996), 
      in Proceedings of the 
      XXVIII International  
      Conference on High Energy Physics,  
      eds. Z.\ Ajduk and A.K.\ Wroblewski, World Scientific, p.53.  
\bibitem{jetrap}
      D0 Collaboration, Phys.\ Rev.\ Lett.\ 
      {\bf 77}, 595 (1996). 
\bibitem{talksjb}
      S.J.\ Brodsky, talk at Workshop on High Energy $e^+ e^-$ Colliders, 
      Brookhaven National Laboratory, May 1996. 
\bibitem{talkfh}
      F.\ Hautmann,   talk at ICHEP96 (Warsaw, July 1996), 
      preprint OITS 613/96,    
      in Proceedings of the
       XXVIII International Conference on High Energy Physics,         
      eds. Z.\ Ajduk and A.K.\ Wroblewski, World Scientific, p.705.    
\bibitem{earlier}
      S.J.\ Brodsky, F.\ Hautmann and D.E.\ Soper, Phys.\ Rev.\ Lett.\ 
      {\bf 78}, 803 (1997). 
\bibitem{gagaphy}
      P.\ Aurenche, G.A.\ Schuler et al., Report 
      on ``$\gamma \, \gamma$ Physics" in Proceedings of the Workshop 
      ``Physics at LEP2", eds. G.\ Altarelli, 
      T.\ Sj{\" o}strand and F.\ Zwirner, CERN 1996-01, Vol.1, p.291.   
\bibitem{onium}
      A.H.\ Mueller, Nucl.\ Phys.\ {\bf B415}, 373 (1994);    
      A.H.\ Mueller and B.\ Patel,
         {\it ibid.} {\bf B425}, 471 (1994).  
\bibitem{bali}
      I.\ Balitskii, Nucl.\ Phys.\ {\bf B463}, 99 (1996). 
\bibitem{bdl}
      J.\ Bartels, A.\ De Roeck and H.\ Lotter, 
      Phys.\ Lett.\ B {\bf 389}, 742 (1996).  
\bibitem{budnev}       
      V.M.\ Budnev, I.F.\ Ginzburg, G.V.\ Meledin and V.G.\ Serbo,
      Phys.\ Rep.\ {\bf 15 C}, 181 (1975). 
\bibitem{lownus}
      F.E.\ Low, Phys.\ Rev.\ D {\bf 12}, 163 (1975); 
      S.\ Nussinov, Phys.\ Rev.\ Lett.\ {\bf 34}, 1286 (1975), 
          Phys.\ Rev.\ D {\bf 14}, 246 (1976); 
      J.F.\ Gunion and D.E.\ Soper, 
          {\it ibid.} {\bf 15}, 2617(1977). 
\bibitem{lcqed}
      J.D.\ Bjorken, J.\ Kogut and 
      D.E.\ Soper, Phys.\ Rev.\ D {\bf 3}, 1382 (1971). 
\bibitem{aligned}
      J.D.\ Bjorken and J.\ Kogut, Phys.\ Rev.\ D {\bf 8}, 1341 (1973). 
\bibitem{frastr}
      L.L.\ Frankfurt and M.\ Strikman, Phys.\ Rep.\ {\bf 160}, 
      235 (1988).  
\bibitem{hef}
      S.\ Catani, M.\ Ciafaloni and F.\ Hautmann, 
         Phys.\ Lett.\ B {\bf 242}, 97 (1990),
         Nucl.\ Phys.\ {\bf B366}, 135 (1991);
      J.C.\ Collins and R.K.\ Ellis,
         {\it ibid.} {\bf B360}, 3 (1991).  
\bibitem{BLM}
      G.P.\ Lepage and P.B.\ Mackenzie,
         Phys.\ Rev.\ D {\bf 48}, 2250 (1993); 
      S.J.\ Brodsky, G.P.\ Lepage and P.B.\ Mackenzie, 
         {\it ibid.} {\bf 28}, 228 (1983).  
\bibitem{bjfutu}
      J.D.\ Bjorken, preprint SLAC-PUB-7341, presented 
      at Snowmass 1996 Summer Study 
      on New Directions for High Energy Physics,   
      e-print archive hep-ph/9610516.
\bibitem{bardiff}
      J.\ Bartels and H.\ Lotter, Phys.\ Lett.\ B {\bf 309}, 400 (1993).  
\bibitem{muni}
      A.H.\ Mueller, Nucl.\ Phys.\ {\bf B437}, 107 (1995). 
\bibitem{sas}
      G.A.\ Schuler and T.\ Sj{\" o}strand, 
          Zeit.\ Phys.\ C {\bf 68}, 607 (1995),
          Phys.\ Lett.\ B {\bf 376}, 193 (1996); 
      M.\ Gl{\" u}ck, E.\ Reya and M.\ Stratmann, 
         Phys.\ Rev.\ D {\bf 51}, 3220 (1995).
\bibitem{regge} 
      P.D.B.\ Collins, {\em An introduction to 
      Regge theory and high energy physics}, Cambridge University 
      Press, Cambridge, 1977.  
\bibitem{lep200angle}
      OPAL Collaboration, contributed paper pa03-007  
      at ICHEP96 (Warsaw, July 1996); 
      J.A.\ Lauber (OPAL),  
      in Proceedings of the
       XXVIII International Conference on High Energy Physics,         
      eds. Z.\ Ajduk and A.K.\ Wroblewski, World Scientific, p.725.      
\bibitem{nlcrep}
      S.\ Kuhlman et al., Physics and Technology of the Next Linear 
      Collider, Snowmass 1996 Report, e-print archive hep-ex/9605011. 

\end{references}
\end{document}